\documentclass[pra,twocolumn,showpacs]{revtex4}
\usepackage{epsfig,amsmath,amssymb,mathtools,amscd}
\usepackage[active]{srcltx}
\def\d{\operatorname{d}}
\def\Exp{\operatorname{Exp}}
\def\bvec#1{\mathbf{#1}}\def\bk{\bvec k}\def\bq{\bvec q}
\def\bh{\bvec h}

\def\bx{\bvec x}

\def\bn{\bvec n}

\def\>{\rangle}
\def\<{\langle}

\def\Tr{\operatorname{Tr}}

\begin{document}

\title{Quantum Cellular Automaton Theory of Light}

 \author{Alessandro   \surname{Bisio}}
 \email[]{alessandro.bisio@unipv.it}
\affiliation{Dipartimento di Fisica dell'Universit\`a di Pavia, via
  Bassi 6, 27100 Pavia} \affiliation{Istituto Nazionale di Fisica
  Nucleare, Gruppo IV, via Bassi 6, 27100 Pavia} 
\author{Giacomo Mauro \surname{D'Ariano}}
 \email[]{dariano@unipv.it}
\affiliation{Dipartimento di Fisica dell'Universit\`a di Pavia, via
  Bassi 6, 27100 Pavia} \affiliation{Istituto Nazionale di Fisica
  Nucleare, Gruppo IV, via Bassi 6, 27100 Pavia} 
\author{Paolo \surname{Perinotti}}
 \email[]{paolo.perinotti@unipv.it}
\affiliation{Dipartimento di Fisica dell'Universit\`a di Pavia, via
  Bassi 6, 27100 Pavia} \affiliation{Istituto Nazionale di Fisica
  Nucleare, Gruppo IV, via Bassi 6, 27100 Pavia} 
\begin{abstract}
  We present a quantum theory of light based on quantum cellular automata (QCA). This approach
  allows us to have a thorough quantum theory of free electrodynamics encompassing an hypothetical
  discrete Planck scale. The theory is particularly relevant because it provides predictions at the
  macroscopic scale that can be experimentally tested. We show how, in the limit of small
  wave-vector $\bk$, the free Maxwell's equations emerge from two Weyl QCAs derived from
  informational principles in Ref.  \cite{d2013derivation}. Within this framework the photon is
  introduced as a composite particle made of a pair of correlated massless Fermions, and the usual
  Bosonic statistics is recovered in the low photon density limit. We derive the main
  phenomenological features of the theory, consisting in dispersive propagation in vacuum, the
  occurrence of a small longitudinal polarization, and a saturation effect originated by the
  Fermionic nature of the photon. We then discuss whether these effects can be experimentally
  tested, and observe that only the dispersive effects are accessible with current technology, from
  observations of arrival times of pulses originated at cosmological distances.
\end{abstract}
\maketitle  

\section{Introduction}

The Quantum Cellular Automaton (QCA) is the quantum version of the popular cellular automaton of von
Neumann \cite{neumann1966theory}. It describes the finite evolution of a discrete set of quantum
systems, each one interacting with a finite number of neighbors via the unitary transformation of
a single step evolution. The idea of a quantum version of a cellular automaton was already contained in the early
work of Feynman \cite{feynman1982simulating}, and later has been object of investigation in the
quantum-information community \cite{schumacher2004reversible,arrighi2011unitarity,gross2012index},
with special enphasis on the so-called Quantum Walks (QW) which decribes the one particle sector of
QCA's with evolution linear in a quantum field
\cite{grossing1988quantum,succi1993lattice,meyer1996quantum,bialynicki1994weyl,ambainis2001one}.

The interest in QCAs is motivated by their potential applications in several fields, like the
statistical mechanics of lattice systems and the quantum computation with microtraps
\cite{cirac2000scalable} and with optical lattices \cite{bloch2004quantum}. Moreover, Quantum Walks
have been used in the design of new quantum algorithms with a computational
speed-up \cite{childs2003exponential,farhi2007quantum}. 

Recently, the idea that QCA could be used to describe a more fundamental discrete Plank scale
dynamics from which the usual Quantum Field Theory emerges
\cite{darianopla,BDTqcaI,d2013derivation}, is gathering increasing attention
\cite{farrelly2014causal,arrighi2013dirac}.  The proposal of modeling Planck scale physics with a
classical automaton on a discrete background first appeared in the work of 't Hooft
\cite{t1990quantization}, and Quantum Walks were considered for the simulation of Lorentz-covariant
differential equations in Refs.
\cite{succi1993lattice,bialynicki1994weyl,meyer1996quantum,PhysRevA.73.054302,Yepez:2006p4406}.

Up to now, most of the interest was focused on the emergence of the Dirac equation for a free
Fermionic field. The choice of cosidering Fermions as the elementary physical systems is motivated
by the idea that the amount of information that can be stored in a finite volume must be finite, as
also suggested by black hole physics \cite{bekenstein1973black,hawking1975particle}.  However, the
question whether a Fermionic QCA could recover the dynamics of a Bosonic field was never addressed
before. Here we will see how free electrodynamics emerges from two Weyl QCAs \cite{d2013derivation}
with Fermionic fields. The dynamical equations resulting in the limit of small wavevector $\bk$ are
the Maxwell's equations. However, for high value of $\bk$ the discreteness of the Planck scale
manifests itself, producing deviations from Maxwell.  Most notably, the QCA dynamics introduces a
$\bk$-dependent speed of light, a feature that was already considered in some approaches to quantum
gravity, and that could be in principle experimentally detected in astrophysical observations
\cite{ellis1992string,lukierski1995classical,Quantidischooft1996,amelino1998tests,amelino2001testable,amelino2001planck,PhysRevLett.88.190403,PhysRevLett.96.051301,ellis2013probes}.

In the present approach the photon turns out to be a composite particle made of a pair of correlated
massless Fermions. This scenario closely resembles the neutrino theory of light of De Broglie
\cite{de1934nouvelle,jordan1935neutrinotheorie,kronig1936relativistically,perkins1972statistics,perkins2002quasibosons}
which suggested that the photon could be composed of a neutrino-antineutrino pair bound by some
interaction.  The failure of the neutrino theory of light was determined by the fact that a
composite particle cannot obey the exact Bosonic commutation relations \cite{pryce1938neutrino}.
However, as it was shown in Ref. \cite{perkins2002quasibosons}, the non-Bosonic terms introduce
negligible contribution at ordinary energy densities.  In our case, as a consequence of the
composite nature of the photon, we have that the number of photons that can occupy a single mode is
bounded. However, as we will see, a saturation effect originated by the Fermionic nature of the
photon is far beyond the current laser technology. 

In Section \ref{sec:quant-cell-autom}, after recalling some basic notions about the QCA,we review
the Weyl automaton of Ref. \cite{d2013derivation}. In Section \ref{s:Maxw} we build a set of
Fermionic bilinear operators, which in Sect. \ref{sec:recov-maxw-dynam} are proved to evolve
according to the Maxwell equations. In Section \ref{sec:photons-as-composite} we will show that the
polarization operators introduced in Sect. \ref{sec:recov-maxw-dynam} can be considered as Bosonic
operators in a low energy density regime.  As a spin-off of this analysis we found a result that
completes the proof, given in Ref.  \cite{PhysRevLett.104.070402}, that the amount of entanglement
quantifies whether pairs of Fermions can be considered as independent Bosons. Section
\ref{sec:phen-analys} presents the phenomenological consequences of the present QCA theory, the most
relevant one being the the appearence of a $\bk$-dependent speed of light. In the same section we
discuss possible experimental tests of such $\bk$-dependence in the astrophysical domain, and we
compare our result with those from Quantum Gravity literature
\cite{ellis1992string,lukierski1995classical,Quantidischooft1996,amelino1998tests,amelino2001testable,amelino2001planck,PhysRevLett.88.190403,PhysRevLett.96.051301,ellis2013probes}.
We conclude with Section \ref{sec:conclusions} where we review the main results and discuss future
developments.

\section{The Weyl automaton: a review}\label{sec:quant-cell-autom}

The basic ingredient of the Maxwell automaton is Weyl's, whis has been derived in Ref. \cite{d2013derivation} from
first principles. Here, we will briefly review the construction for completeness. 

A QCA represents the evolution of a numerable set $G$ of cells $g\in G$, each one containing an
array of Fermionic local modes. The evolution occurs in discrete identical steps, and in each one
every cell interacts with a the others. The Weyl automaton is derived from the following principles:
unitarity, linearity, locality, homogeneity, transitivity, and isotropy. Unitarity means just that
each step is a unitary evolution. Linearity means that the unitary evolution is linear in the field.
Locality means that at each step every cell interacts with a finite number of others. We call cells
interacting in one step {\em neighbors}. The neighboring notion also naturally defines a graph over
the automaton, with $g$ as vertices and the neighboring couples as edges. Homogeneity means both
that all steps are the same, all cells are identical systems, and the set of interactions with
neigbours is the same for each cell, hence also the number of neigbours, and the dimension of the
cell field array, which we will denote by $s>0$. We will denote by $A$ the matrix representing the
linear unitary step. Transitivity means that every two cells are connected by a path of neighbours.
Isotropy means that the neighboring relation is symmetric, and there exists a group of automorphisms
for the graph for which the automaton itself is covariant. Homogeneity, transitivity, and isotropy
together imply that $G$ is a group, and the graph is a Cayley graph $\Gamma(G,S_+)$ where
$G=\<S_+|R\>$ is a presentation of $G$ with generator set $S_+$ and relator set $R$. The set of
neighboring cells is then given by $S:=S_+\cup S_-$ where $S_-$ is the set of the inverse
generators. Linearity, locality, and homogeneity imply that each step can be described in terms of
transition matrices $A_h\in\rm{M}(\mathbb{C},s)$ for each $h\in S$, and then the step is described
mathematically as follows
\begin{align}
  \label{eq:automagraph}
  \psi_g(t+1) = \sum_{h\in S}A_h\psi_{hg}(t)
\end{align}
where $\psi_g(t)$ is the $s$-array of field operators at $g$ at step $t$. Therefore, upon denoting
by $T_g$ $g\in G$ the unitary representation of $G$ on $\ell^2(G)$, $T_g|f\>:=|gf\>$, for $f\in G$,
$A$ is a unitary operator on $\ell^2(G)\otimes\mathbb{C}^s$ of the form
\begin{align}
A:=\sum_{h\in S}T_h\otimes A_h.
\label{eq:walk}
\end{align}
Covariance of the isotropy property means precisely that the group $L$ of automorphisms of the graph
is a transitive permutation group of $S_+$, and there exists a (generally projective) unitary
representation $U_l$ $l\in L$ of $L$ such that
\begin{align}
A=\sum_{h\in S}T_{lh}\otimes U_l A_{h}U_l^\dag,\qquad \forall l\in L.
\label{eq:iso}
\end{align}

In Ref.~\cite{d2013derivation} attention was restricted to group $G$ quasi-isometrically embeddable
in an Euclidean space, which is then {\em virtually Abelian} \cite{Cornulier07}, namely it has an
Abelian subgroup $G'\subset G$ of finite index, namely with a finite number of cosets. Then it can
be shown the automaton is equivalent to another one with group $G'$ and dimension $s'$ multiple of
$s$. We further assume that the representation of the isotropy group $L$ induced by the embedding is
orthogonal, which implies that the graph neighborhood is embedded in a sphere. We call such a
property {\em orthogonal isotropy}.

For $s=1$ the automaton is trivial, namely $A=I$. For $s=2$ and for Euclidean space $\mathbb R^3$
one has $G=\mathbb Z^3$, and the Cayley graphs satisfying orthogonal isotropy are the Bravais
lattices. The only lattice that has a nontrivial set of transition matrices giving a unitary
automaton is the BCC lattice. We will label the group element as vectors $\bx\in\mathbb{Z}^3$, and
use the customary additive notation for the group composition, whereas the unitary representation of
$\mathbb{Z}^3$ is expressed as follows
\begin{equation}
T_{\bvec z}|\bvec x\>=|\bvec z+\bvec x\>.
\end{equation}
Being the group Abelian, we can Fourier transform, and the operator $A$ can be easily block-diagonalized
in the $\bk$ representation as follows
\begin{align}
  \label{eq:weylautomata} 
  A = \int_B\operatorname d^3 \! \bk  \,  |{\bk}\>\< {\bk}| \otimes A_{\bk}
\end{align}
with $A_\bk:=\sum_{\bvec h\in S}\bvec k\,e^{-i\bvec k\cdot\bvec h}A_\bh$ unitary for every $\bk\in
B$, and the vectors $|{\bk}\>$ given by
\begin{equation}
|\bk\>:=\frac1{\sqrt{2\pi}^3}\sum_{\bx\in G}e^{i\bk\cdot\bx}|\bx\>,
\end{equation}
is a Dirac-notation for the direct integral over $\bk$, and the domain $B$ is the first Brillouin zone of
the BCC. There are only two QCAs, with unitary matrices
\begin{equation}\label{eq:weylautomata2} 
A^{\pm}_{\bk} := d^{\pm}_{\bk} I+\tilde{\bn}^{\pm}_{\bk}\cdot\boldsymbol{\sigma}
=\exp[-i\bvec{n}^{\pm}_{\bk} \cdot \boldsymbol{\sigma}],
\end{equation}
where
\begin{align}
&\tilde{\bn}^{\pm}_{\bk} :=
\begin{pmatrix}
s_x c_y c_z \mp c_x s_y s_z\\
\mp c_x s_y c_z - s_x c_y s_z\\
c_x c_y s_z \mp s_x s_y c_z
\end{pmatrix}\!\!,\,
{\bn}^{\pm}_{\bk}:=\frac{\lambda^{\pm}_{\bk}\tilde{\bn}^{\pm}_{\bk}}{\sin\lambda^{\pm}_{\bk}},\nonumber\\
&\d^{\pm}_{\bk} :=  (c_x c_y c_z \pm s_x s_y s_z ),\;
\lambda^{\pm}_{\bk}:=\arccos(d^{\pm}_{\bk}),\nonumber
\end{align}
and
\begin{equation}
c_\alpha := \cos({k}_\alpha/\sqrt{3}),\;s_\alpha:= \sin({k}_\alpha/\sqrt{3}),\;\alpha = x,y,z.\nonumber
\end{equation}
The matrices $A_{\bk}^\pm$ in Eq. \eqref{eq:weylautomata2} describe the evolution of a two-component
Fermionic field,
\begin{align}
  \label{eq:automa1}
  {\psi} ({\bk},t+1) = 
A_{\bk}^\pm {\psi} ({\bk},t),
\quad
  {\psi} ({\bk},t) : =
\begin{pmatrix}   
 {\psi}_R ({\bk},t)\\
 {\psi}_L ({\bk},t)
  \end{pmatrix}.
  \end{align}
  The adimensional framework of the automaton corresponds to measure everything in Planck units. In
  such a case the limit $|{\bk}|\ll 1$ corresponds to the relativistic limit, where on has
\begin{equation}
\bn^{\pm}({\bk})\sim\tfrac{{\bk}}{\sqrt{3}},\quad A^{\pm}_{\bk}\sim\exp[-i\tfrac{{\bk}}{\sqrt{3}} \cdot\boldsymbol{\sigma}],
\end{equation}
corresponding to the Weyl's evolution, with $\tfrac{{\bk}}{\sqrt{3}}$ playing the role of momentum.

\section{The Maxwell automaton}\label{s:Maxw}
In order to build the Maxwell dynamics, we need to consider two different Weyl QCAs the first one
acting on a Fermionic field $\psi(\bk)$ by matrix $A_\bk$ as in Eq.  (\ref{eq:automa1}), and the
second one acting on the field $\varphi(\bk)$ by the complex conjugate matrix $A_\bk^*=\sigma_y
A_\bk\sigma_y$, i.e.
\begin{align}
  \label{eq:automa2}
  {\varphi} (\bk,t+1) = A_\bk^*{\varphi} (\bk,t), \quad
  {\varphi} (\bk,t)  =
  \begin{pmatrix}
    {\varphi}_R (\bk,t)\\
    {\varphi}_L (\bk,t)
  \end{pmatrix}.
\end{align}
The matrix $A_\bk$ can be either one of the Weyl matrices $A^\pm_{\bk}$, and the whole derivation is
independent of the choice.

The Fermionic fields ${\varphi}$ and ${\psi}$ are independent and obey the following
anti-commutation relations
  \begin{align}
&[\psi_i(\bk),\psi_j(\bk')  ]_+ =
[\varphi_i(\bk),\varphi_j(\bk')  ]_+ =\nonumber\\
&[\varphi_i(\bk),\psi_j(\bk')  ]_+=
[\varphi_i(\bk),\psi^\dagger_j(\bk')  ]_+
=0      
\nonumber \\
&[\psi_i(\bk),\psi^\dagger_j(\bk')  ]_+ =
[\varphi_i(\bk),\varphi^\dagger_j(\bk')  ]_+
=
\delta_B(\bk-\bk')\delta_{i,j} \nonumber \\ 
 &i,j = R,L \qquad   \bk,\bk' \in B,
\label{eq:commutationrel}
\end{align}
where $\delta_B(\bk)$ is the 3d Dirac's comb delta-distribution (which repeats periodically with
$\mathbb R^3$ tasselated into Brillouin zones).

Given now two arbitrary fields ${\eta}(\bk)$ and ${\theta}(\bk)$ we
define the following bilinear function
\begin{align}
  \label{eq:gimu}
\!\!  G_f^{\mu}(\eta,\theta,\bk) := \!\!
\int\!\!\frac{\d\bq}{(2\pi)^3} f_{\bk}(\bq)
{{\eta}}^T
\left(\tfrac{\bk}{2}-\bq\right)
\sigma^{\mu}
{\theta}
 \left(\tfrac{\bk}{2}+\bq\right) 
\end{align}
where $\sigma^0:= I $, $\sigma^1:= \sigma^x$, $\sigma^2:= \sigma^y $, $\sigma^3:=\sigma^z$ and
$\int\frac{\d\bq}{(2\pi)^3} |f_{\bk}(\bq)|^2 =1, \forall \bk$. In the following we will also treat
the vector part $\boldsymbol\sigma:=(\sigma^1,\sigma^2,\sigma^3)$ of the four-vector $\sigma^\mu$
separately.  This allows us to define the following operators
\begin{align}
  \label{eq:bilinears}
  F^{\mu}(\bk) :=G_f^\mu(\varphi,\psi,\bk)
\end{align}

In the following sections we study the evolution of the bilinear functions $F^{\mu}(\bk)$ and their
commutation relations and show that, in the relativistic limit and for small particle densities the
quantum Maxwell equations are recovered for both choices of $A_\bk=A^\pm_\bk$.

\section{The Maxwell dynamics}\label{sec:recov-maxw-dynam}

In the following we will use the short notations 
\begin{equation}\label{eq:notaz}
 [Z\eta](\bk):=Z_{\bk}\eta(\bk),\quad [ZW]_\bk:=Z_{\bk}W_\bk,
\end{equation}
for $\eta$ a field and $Z$ and $W$ matrices.  If the fields $\psi$ and $\varphi$ evolve according to
Eqs. \eqref{eq:automa1} and \eqref{eq:automa2}, then the evolution of the bilinear functions
$F^{\mu}(\bk)$ introduced in Eq. \eqref{eq:bilinears} obeys the following equation
\begin{align}
\label{eq:exactevolution}
   &F^{\mu}(\bk,t) = G_f^\mu([{A^*}^t\varphi],[A^t\psi],\bk),
\end{align}
where we used the notation in (\ref{eq:notaz}). Now, let us define
\begin{align}
 &\tilde F^\mu(\bk,t):= G_f^\mu([{U^{\bk,t}}^*\varphi],[U^{\bk,t}\psi],\bk), \nonumber\\ 
 &U^{\bk,t}_\bq :=A^{-t}_{\tfrac{\bk}2} A^t_\bq,
\label{eq:defU}
\end{align}
where we remind that $[{U^{\bk,t}}^*\varphi](\bq):={U_\bq^{\bk,t}}^*\varphi(\bq)$.  Clearly, one has
$[A^t\eta]=[A_{\frac{\bk}{2}}^tU^{\bk,t}\eta]$. We now need the identity
\begin{align}
&\exp (-\tfrac{i}{2}\bvec{v}\cdot \boldsymbol{\sigma}) 
\boldsymbol{\sigma} \exp (\tfrac{i}{2}\bvec{v} \cdot \boldsymbol{\sigma}) =
\Exp(-i\bvec{v} \cdot \bvec{J}) \boldsymbol{\sigma},\nonumber\\
&\exp (-\tfrac{i}{2}\bvec{v}\cdot \boldsymbol{\sigma}) 
\sigma^0 \exp (\tfrac{i}{2}\bvec{v} \cdot \boldsymbol{\sigma}) =\sigma^0,
\end{align}
where the matrix $\Exp(-i\bvec{v}\cdot\bvec{J})$ acts on $\boldsymbol{\sigma}$ regarded as a vector,
and $\bvec J=(J_x, J_y,J_z)$ is the vector of angular momentum operators. We can then recast
Eq.~\eqref{eq:exactevolution} in terms of the following functions
\begin{align}
\bvec{F}(\bk,t) &:= 
(
F^{1}(\bk,t), F^{2}(\bk,t), F^{3}(\bk,t)
) ^T, \label{eq:ftilde}
\end{align}
and $\tilde{\bvec{F}}(\bk,t)$ similarly defined, obtaining
\begin{align}
\label{eq:evolutionwithrotation}
    & F^{0}(\bk,t) =
\tilde{F}^{0}(\bk,t), \nonumber\\
& \bvec{F}(\bk,t)  =
\Exp\left(-2i {\bvec{n}}_{\tfrac{\bk}{2}} \cdot
      \bvec{J}t\right)
\tilde{\bvec{F}}(\bk,t).  
\end{align}
If we  assume that
\begin{align}
  \label{eq:approxf}
 \int_{|\bq| \geq \bar{q}(\bk)}\frac{\d\bq}{(2\pi)^3} |f_{\bk}(\bq)|^2
 \ll 1 \quad\mbox{for}\  \bar{q}(\bk) \ll |\bk|,  
\end{align}
by taking the Taylor expansion of ${\bvec{n}}_{\tfrac{\bk}{2}+\bq}$
with respect to $\bq$ we can make the approximation
\begin{align}
        {U}^{\bk,t}_{\tfrac{\bk}2\pm\bq}&\simeq \exp\left(i
          {\bn}_{\tfrac{\bk}{2}} \cdot \boldsymbol{\sigma}t\right)
        \exp\left[-i\left({\bn}_{\tfrac{\bk}{2}}
            \pm\bvec{l}_{\bk,\bq}\right) \cdot
          \boldsymbol{\sigma}t\right] \nonumber \\
&
\simeq
\exp \left(\pm
i c_{\bk,\bq} 
\frac{{\bvec{n}}_{\frac{\bk}{2}}}{|{\bvec{n}}_{\frac{\bk}{2}}|} \cdot \boldsymbol{\sigma}
t\right)+ O \big(  \tfrac{\bar{q}(\bk)}{|\bvec{n}_{\frac{\bk}{2}}|}
\big) 	\label{eq:approxU0}
,
\end{align}
where $\bvec{l}_{\bk, \bq} := J_{{\bvec{n}}}\left(\frac{\bk}{2}\right)\bq $ and
$J_{{\bvec{n}}}\left(\frac{\bk}{2}\right)$ denotes the Jacobian matrix of the function
$\bvec{n}_\bk$ evaluated at $\frac{\bk}{2}$ and $c_{\bk,\bq} :=
\frac{{\bvec{n}}_{\frac{\bk}{2}}}{|{\bvec{n}}_{\frac{\bk}{2}}|} \cdot
\bvec{l}_{\bk, \bq}$ (the proof of Eq. \ref{eq:approxU0} is given
in Appendix \ref{sec:proof-eq.-eqref}).
By introducing the transverse field operators
\begin{align}
  \label{eq:transverse}
  \begin{split}
 \tilde{\bvec{F}}_T(\bk,t) :=\tilde{\bvec{F}}(\bk,t)- 
\left(\frac{\bvec{n}_{\frac{\bk}{2}}}{|\bvec{n}_{\frac{\bk}{2}}|} \cdot
  \tilde{\bvec{F}}(\bk,t) \right)
\frac{\bvec{n}_{\frac{\bk}{2}}}{|\bvec{n}_{\frac{\bk}{2}}|} \\
\bvec{F}_T(\bk,t) := \bvec{F}(\bk,t)  - 
\left(\frac{\bvec{n}_{\frac{\bk}{2}}}{|\bvec{n}_{\frac{\bk}{2}}|} \cdot
{\bvec{F}}(\bk,t) \right)
\frac{\bvec{n}_{\frac{\bk}{2}}}{|\bvec{n}_{\frac{\bk}{2}}|}.
  \end{split}
\end{align} 
and using Eq.  \eqref{eq:approxU0} into Eq. \eqref{eq:ftilde} we
get
(see Appendix \ref{sec:proof-eq.-eqref-1})
\begin{align}
  \label{eq:transverse2}
  \begin{split}
    \tilde{\bvec{F}}_T(\bk,t) =  
  {\bvec{F}}_T(\bk) 
+
 O \big(  \tfrac{\bar{q}(\bk)}{|\bvec{n}_{\frac{\bk}{2}}|} \big).
  \end{split}
\end{align}
Finally, combining Eq. \eqref{eq:transverse2} 
with Eq. \eqref{eq:evolutionwithrotation} we obtain a closed
expression for the time evolution of the operator 
${\bvec{F}_T}(\bk)$,
\begin{align}
  \begin{split}
    \label{eq:maxwell}
  \bvec{F}_T(\bk,t)  =
\exp\left[\left(2 \bvec{n}_{\tfrac{\bk}{2}} \cdot
      \bvec{J}\right)t\right]
{\bvec{F}_T}(\bk) +\Lambda(\bk,t),  
  \end{split}
\end{align}
where $\|\Lambda(\bk,t)\|= O \big( \tfrac{\bar{q}(\bk)}{|\bvec{n}_{\frac{\bk}{2}}|} \big)$.  Taking
the time derivative in Eq. \eqref{eq:maxwell} and reminding the definition \eqref{eq:transverse} we
obtain
\begin{align}
  \begin{split} \label{eq:maxwell2}
    &\partial_t\bvec{F}_T(\bk,t) =
2\bvec{n}_{\frac{\bk}{2}} \times \bvec{F}_T(\bk,t)+
\partial_t \Lambda(\bk,t)\\
&2\bvec{n}_{\frac{\bk}{2}} \cdot \bvec{F}_T(\bk,t) = 0,
  \end{split}
\end{align}
where $\|\partial_t \Lambda(\bk,t)\|=O \big(
\tfrac{\bar{q}(\bk)}{|\bvec{n}_{\frac{\bk}{2}}|} \big)$ (see Appendix \ref{sec:proof-eq.-eqref-1}).

\begin{figure}[t]
  \begin{center}
    \includegraphics[width=8cm ]{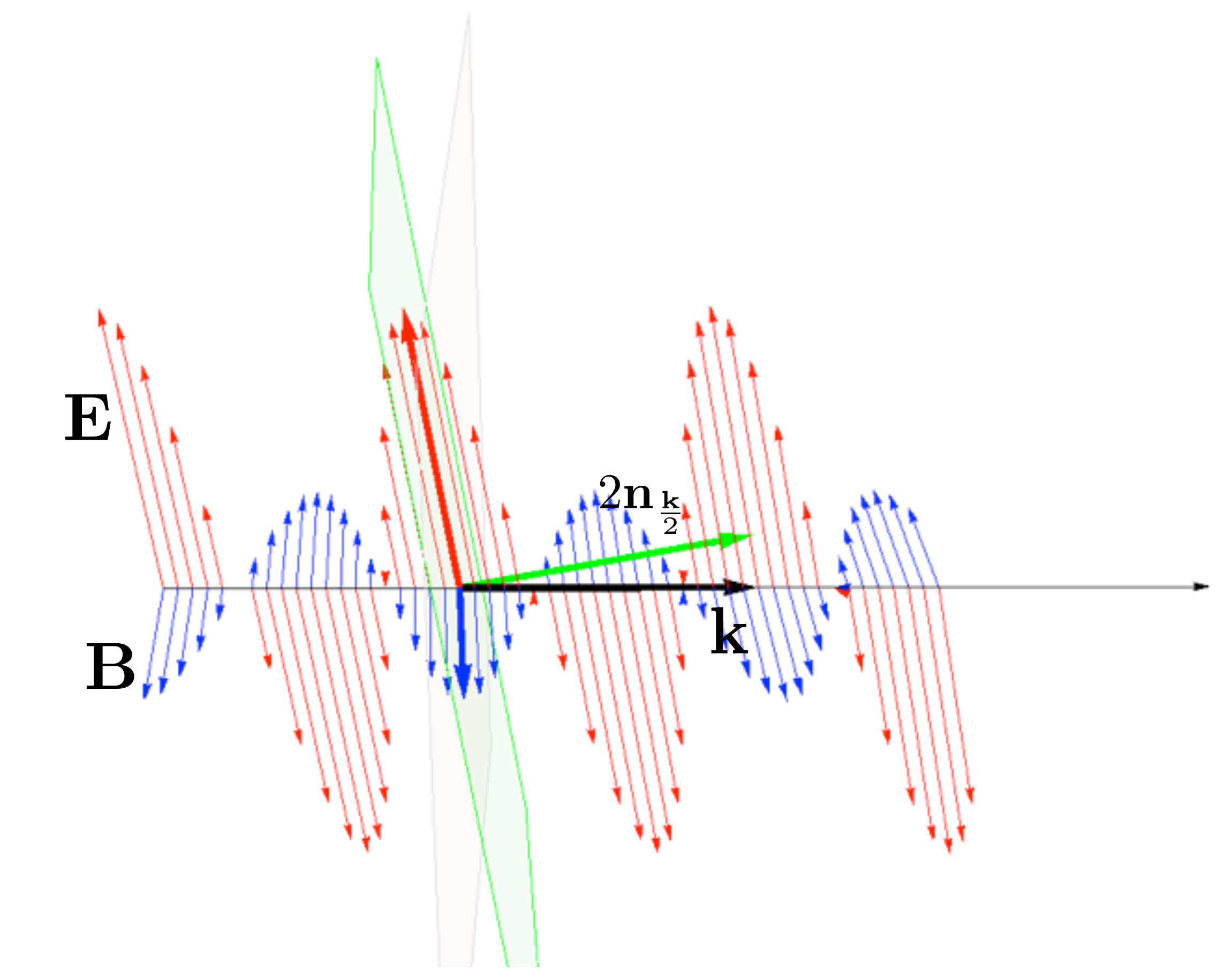}
    \caption{(colors online) A
      rectilinear polarized electromagnetic wave. We notice that the
      polarization plane (in green) is sligtly tilted with respect the
      plane orthogonal to $\bk$ (in gray).}
    \label{fig:elmwave}
  \end{center}
\end{figure}

Let now $\bvec{E}$ and
$\bvec{B}$ be two Hermitian operators defined by the relation
\begin{align}
  \label{eq:electric and magnetic field}
  &\bvec{E}:=|{\bn}_{\tfrac\bk2}|(\bvec{F}_T+\bvec{F}_T^\dag),\quad\bvec{B}:=i|{\bn}_{\tfrac\bk2}|(\bvec{F}_T^\dag-\bvec{F}_T),\nonumber\\
  &2|{\bn}_{\tfrac\bk2}|\bvec{F}_T=\bvec{E} + i \bvec{B}.
\end{align}
We now show that
in the limit of small wavevectors $\bk$
and by interpreting
$ \bvec{E}$ and $ \bvec{B} $
as the electric and magnetic field
 the usual vacuum Maxwell's equations can be recovered.
For $|\bk| \ll 1$  one has $2\bvec{n}_{\frac{\bk}{2}} \simeq \bk/\sqrt3$, and
Eq. \eqref{eq:maxwell2} becomes
\begin{align}
  \begin{split} \label{eq:maxwell3}
    &\partial_t\bvec{F}_T(\bk,t) =
\frac{\bk}{\sqrt3} \times \bvec{F}_T(\bk,t)
\\
&\bk\cdot \bvec{F}_T(\bk,t) = 0
  \end{split} \; .
\end{align}
As in Ref. \cite{d2013derivation}, we recover physical dimensions from the previous adimensional
equations using Planck units, taking $c:=l_P/t_P$, time measured in Planck times $t\to t*t_P$, 
and lengths measured in Planck lenghts as $x\to x*\sqrt{3}l_P$, the $\sqrt{3}l_P$ corresponding to
the distance between neighboring cells. Then Eq. \eqref{eq:maxwell3} becomes
\begin{align}
  \begin{split} \label{eq:maxwellposition}
    &\partial_t\bvec{F}_T(\bvec{x},t) =
-ic\nabla\times \bvec{F}_T(\bvec{x},t)
\\
&\nabla \cdot \bvec{F}_T(\bvec{x},t) = 0
  \end{split} 
\end{align}
which in terms of $\bvec{E}$ and $\bvec{B}$ become the vacuum Maxwell's equations
\begin{align}
      \label{eq:maxwellstandard}
  \begin{array}{lcl}
    \nabla \cdot \bvec{E}=0    & &\nabla \cdot \bvec{B} =0\\
    \partial_t \bvec{E} = c\nabla \times \bvec{B}  && \partial_t \bvec{B} = -c\nabla \times \bvec{E} \;\;.
  \end{array}
\end{align}
Introducing the polarization vectors $\bvec{u}_\bk^1$ and $\bvec{u}_\bk^2$ satisfying
\begin{equation}
\bvec{u}^i_\bk \cdot \bn_{\bk} =\bvec u^1_\bk\cdot\bvec u^2_\bk= 0,\ |\bvec u^i_\bk|=1,\ 
(\bvec u^1_\bk\times\bvec u^2_\bk)\cdot\bn_\bk>0,
\end{equation}
we can now interpret the following operators
\begin{align}
  \gamma^i(\bk) &:= \bvec{u}^i_\bk\cdot\bvec{F}(\bk,0),\quad i=1,2,
  \label{eq:polarization}
\end{align}
as the two polarization operators of the field.  In the light of this analysis, one can conclude
that the automaton discrete evolution leads to modified Maxwell's equations in the form of Eqs.
\eqref{eq:maxwell2}, with the electromagnetic field rotating around $\bn_{\tfrac{\bk}{2}}$ instead
of $\bk$.
Moreover, since in this framework the photon is a composite particle, the internal dynamics of the
consitutent Fermions is responsible for an additional term $O \big(
\tfrac{\bar{q}(\bk)}{|\bvec{n}_{\frac{\bk}{2}}|} \big)$.  As a consequence of this distorsion, one
can immediately see that the electric and magnetic fields are no longer exactly transverse to the
wave vector but we have the appearence of a longitudinal component of the polarization (see Fig.
\ref{fig:elmwave}).  In Section \ref{sec:phen-analys} we discuss the new phenomenology that emerges
from Eqs. \eqref{eq:maxwell2}.

\section{Photons as composite Bosons}\label{sec:photons-as-composite}
                         
In the previous section we proved that the operators defined
in Eq. \eqref{eq:electric and magnetic field} dynamically evolve
according to the free Maxwell's equation. However,
 in order to interpret
$\bvec{E}(\bk)$ and $\bvec{B}(\bk)$ as the electric and magnetic fields we need
to show that they obey the correct commutation relation.
The aim of this paragraph is to show that, in a regime of low energy
density, the polarization operators defined in Eq. \eqref{eq:polarization}
actually behave as independent Bosonic modes.

In order to avoid the technicalities of the continuum we now 
suppose to confine the system in finite volume $\mathcal{V}$.
The finiteness of the volume introduces a discretization of the 
momentum space and the operators $\psi(\bk)$, $\varphi(\bk)$, 
obey Eq.~\eqref{eq:commutationrel} where the periodic Dirac delta is replaced by the Kronecker delta. 
All the integrals over the Brillouin zone are then
replaced by sums, and the polarization operators  of Eq. \eqref{eq:polarization} become
\begin{equation}
  \gamma^i(\bk) := 
\sum_{\bq}
f_{\bk}(\bq)
{\varphi}^T \left(\tfrac{\bk}{2}-\bq\right)
(
\bvec{u}^i_{\tfrac{\bk}2}
\cdot
\boldsymbol{\sigma}
)
{\psi} \left(\tfrac{\bk}{2}+\bq\right).
\end{equation}
These operators can be simply expressed in terms of the functions $\gamma_{\alpha,\beta}(\bvec{k})$ defined as follows
\begin{align}
  \gamma_{\alpha,\beta}(\bvec{k}) := \sum_{\bvec{q}}
   {f}_{\bk}(\bvec{q}) 
\varphi_\alpha
\left(\tfrac{\bk}{2}-\bq\right)
\psi_\beta
\left(\tfrac{\bk}{2}+\bq\right),
\nonumber \\
\alpha, \beta = R,L. 
  \label{eq:basicobjects}
\end{align}
Since the polarisation operators $\gamma^i(\bk)$ are linear
combinations of $\gamma_{\alpha,\beta}(\bk)$, it is useful to compute
the commutation relations of the latter. We have
\begin{align}
  \label{eq:basiccommutation}
  &[\gamma_{\alpha,\beta}(\bvec{k}) ,\gamma_{\alpha',\beta'}(\bvec{k}') ]_{-} = 0,\nonumber\\
  &[\gamma_{\alpha,\beta}(\bvec{k})
  ,\gamma^\dagger_{\alpha',\beta'}(\bvec{k}') ]_{-} =
  \delta_{\alpha,\alpha'} \delta_{\beta,\beta'}
  \delta_{\bvec{k},\bvec{k}'} -\Delta_{\alpha,\alpha',\beta,\beta',\bk,\bk'},\nonumber\\
  &\Delta_{\alpha,\alpha',\beta,\beta',\bk,\bk'}:=\left( \delta_{\alpha,\alpha'}H^+_{\psi, \beta', \beta,\bvec{k}',\bvec{k}}+ \delta_{\beta,\beta'}H^-_{\varphi, \alpha', \alpha,\bvec{k}',\bvec{k}}\right), \nonumber\\
  &H^\pm_{\eta, \alpha', \alpha,\bvec{k}',\bvec{k}} := \sum_{\bvec{q}}{f}_{\bk}(\bvec{q}) {f}_{\bk'}^*(\tfrac{\bvec{k}' - \bvec{k}}2+\bvec{q})\nonumber\\
  &\qquad\times\eta_{\alpha'}^\dagger \left(\tfrac{ 2\bvec{k}' -
    \bvec{k}}2\pm\bvec{q}\right) \eta_{\alpha} \left(
    \tfrac{\bvec{k}}2\pm{\bvec{q}}\right).
\end{align}
Then the operators 
$\gamma_{\alpha,\beta}$ fail to be Bosonic annihilation operators
because of 
the apperance of the operator
$\Delta_{\alpha,\alpha', \beta, \beta',\bvec{k},\bvec{k}'}$ in the commutation relation
 \eqref{eq:basiccommutation}.
However, if we restrict to the subset $\mathcal{S}$ of states such that
$\Tr[\rho H^-_{\varphi, \beta', \beta,\bvec{k}',\bvec{k}}] \simeq 0$ and
$\Tr[\rho H^+_{\psi, \alpha',  \alpha,\bvec{k}',\bvec{k}}] \simeq 0$ for all $\rho \in
\mathcal{S}$, we could make the approximation 
$[\gamma_{\alpha,\beta}(\bvec{k}) ,\gamma^\dagger_{\alpha',\beta'}(\bvec{k}')
  ]_{-} \simeq 
\delta_{\alpha,\alpha'} \delta_{\beta,\beta'}
\delta_{\bvec{k},\bvec{k}'} $.
If we consider the modulus of the expectation value
of the operators
$H^\pm_{\eta, \beta', \beta,\bvec{k}',\bvec{k}} $
we have
\begin{align}
  \label{eq:boundfordelta}
&|\< H^\pm_{\eta, \beta', \beta,\bvec{k}',\bvec{k}} \>| 
 \leq\sum_{\bvec{q}}
   \left| {f}_{\bk}(\bvec{q}) \right |
 \left| {f}_{\bk'}^*(\tfrac{\bvec{k}' - \bvec{k}}2+\bvec{q}) \right |\nonumber\\
 &\quad\times
\left|\left\<  \eta_{\beta'}^\dagger \left( \tfrac{2\bvec{k}' -
    \bvec{k}}2\pm\bvec{q}\right)
\eta_{\beta} \left(\tfrac{ \bvec{k}}2\pm\bvec{q}\right )
\right \> \right | \leq\nonumber\\
&\qquad\sqrt{
\<
\Gamma^\pm_{\eta,\beta,\bvec{k}}
\>
\<
\Gamma^\pm_{\eta,\beta',\bvec{k}'}
\>
},\\
&\Gamma^\pm_{\eta,\beta,\bvec{k}} =
\sum_{\bvec{q}}
   \left|{f}_{\bk}(\bvec{q}) \right |^2
\eta^\dagger_{\beta} \left(\tfrac{ \bvec{k}}2\pm\bvec{q}\right )
\eta_{\beta} \left(\tfrac{ \bvec{k}}2\pm\bvec{q}\right ),
\end{align}  
where we repeatedly applied the Schwartz inequality. 

The operators
 $\Gamma^-_{\varphi,\beta,\bvec{k}}$ and 
$\Gamma^+_{\psi, \alpha,\bvec{k}}$ can be interpreted as 
number operators ``shaped'' by the probability distribution
$|{f}_{\bk}(\bvec{q})|^2$.
If we suppose
$| {f}_\bk(\bvec{q})|^2$ to be a constant function
over a region $\Omega_\bk$ which contains $N_\bk$ modes,
i.e. $|{f}_\bk(\bvec{q})|^2 = \tfrac{1}{N_\bk}$ if $\bvec{q}\in \Omega_\bk$
and $|{f}_\bk(\bvec{q})|^2 = 0 $ if $\bvec{q} \not\in \Omega_\bk$, we have
\begin{align*}
 \left\<
\Gamma^+_{\psi, \alpha,\bvec{k}}
\right\> = 
\frac{1}{N_\bk}\sum_{\bvec{q}\in\Omega_\bk}
\left\<
\psi^\dagger_{\alpha} \left( \tfrac{\bvec{k}}2+\bvec{q}\right )
\psi_{\alpha} \left(\tfrac{ \bvec{k}}2+\bvec{q}\right )
\right\> = \frac{M_{\psi,\alpha,\bvec{k}}}{N_\bk}
\end{align*}
where we denoted with $M_{\psi,\alpha,\bvec{k}}$ the number of
$\psi_{\alpha}$ Fermions in the region $\Omega_k $ (clearly the same
result applies to $\Gamma^-_{\varphi, \beta,\bvec{k}}$).  Then, if we
consider states $\rho$ such that $M_{\xi,\chi,\bvec{k}}/ N_\bk \leq
\varepsilon$ for all $\xi_{\chi}$ and $\bvec{k}$ and for $\varepsilon \ll 1$
we can safely assume $[\gamma_{\alpha,\beta}(\bvec{k})
,\gamma^\dagger_{\alpha',\beta'}(\bvec{k}') ]_{-} =
\delta_{\alpha,\alpha'} \delta_{\beta,\beta'}
\delta_{\bvec{k},\bvec{k}'} $ in Eq. \eqref{eq:basiccommutation} which
after an easy calculation gives
\begin{align}
  \label{eq:commutationpolarization}
  [\gamma^i (\bvec{k}),{\gamma^j}^\dag (\bvec{k}')]_-  =
\delta_{i,j} \delta_{\bvec{k},\bvec{k}'}\quad i = 0,1,2,3.
\end{align}
In Eq. \eqref{eq:commutationpolarization}, besides the previously defined transverse polarizations
$\gamma^1 (\bvec{k}) $ and
$\gamma^2 (\bvec{k}) $,
 we considered also the ``longitudinal''
polarization operator
$\gamma^3 (\bvec{k}) := \sum_{\bvec{q}}
   f_\bk(\bvec{q}) 
{\varphi}^T
\left(  \tfrac{\bvec{k}}2 -\bvec{q} \right)
(
\bvec{e}_{\tfrac{\bk}2} \cdot
\boldsymbol{\sigma}
)
{\psi} 
\left(\tfrac{
\bvec{k}}2 +\bvec{q}
\right)$, where $\bvec e_{\bk}:=\bn_\bk/|\bn_\bk|$,
and the ``timelike'' polarization operator
$\gamma^0 (\bvec{k}) := \sum_{\bvec{q}}
   {f}_\bk(\bvec{q}) 
{\varphi}^T
\left( \tfrac{ \bvec{k}}2 - \bvec{q} \right)
I
{\psi} 
\left(
\tfrac{\bvec{k}}2+\bvec{q}
\right)$.

This result tells us that, as far as we restrict ourselves to states in
$\mathcal{S}$ we are allowed to interpret the operators 
$\gamma^i(\bvec{k})$ as $4$ independent Bosonic field modes and then
to interpret $\bvec{E}$ and $\bvec{B}$ defined in
Eq. \eqref{eq:electric and magnetic field}
as the electric and the magnetic field operators.
This fact together with the evolution given by Eq. \eqref{eq:maxwell3}
proves that we realized a consistent model of quantum electrodynamics
in which the photons are composite particles made by correlated
Fermions whose evolution is described by a cellular automaton.

\subsection{Composite Bosons and entanglement}\label{sec:comp-bosons-entangl}

The results that we had in this section are in agreement with the recent works
\cite{combescot2001new, rombouts2002maximum, avancini2003compositeness,combescot2003n} which studied
the conditions under which a pair of Fermionic fields can be considered as a Boson.  In Refs.
\cite{PhysRevA.71.034306, PhysRevLett.104.070402} it was shown that a sufficient condition is that
the two Fermionic fields $\psi,\phi$ are sufficiently entangled. More precisely, for a composite
Boson $c := \sum_{i} f(i) \psi_i \phi_i $, $\sum_{i} |f(i)|^2=1$ one has
\begin{equation}
[c,c^\dag] = 1- (\Gamma_\psi + \Gamma_{\phi}),
\end{equation}
where 
\begin{equation}
\Gamma_\psi = \sum_{i} |f(i)|^2 \psi^\dag_i \psi_i,\quad\Gamma_\phi = \sum_{i} |f(i)|^2 \phi^\dag_i \phi_i,
\end{equation}
and in Ref.~\cite{PhysRevLett.104.070402} it was shown that the following bound holds
\begin{equation}
\forall N \geq 1,\quad NP \geq \<{N}|\Gamma_\psi |{N} \>\geq P,
\end{equation}
and the same holds for $\Gamma_\phi$, where $P = \sum_{i=1}^N |f(i)|^4$ is the purity of the reduced
state of a single particle and $|{N}\> = \tfrac{1}{\sqrt{N!}}  \chi_N(c^\dag)^N |{0}\>$ ($\chi_N$ is
a normalization constant).  From this result, the authors of Ref.  \cite{PhysRevLett.104.070402}
concluded that, as far as $P,NP \approx 0$, $c$ and $c^\dag$ can be safely considered as a Bosonic
annihilation/creation pair. Our criterion, which restricts the state $\rho$ to satisfy $\Tr[\rho
\Gamma_\psi],\Tr[\rho\Gamma_\phi] \leq \varepsilon$ in this simplified scenario, gives the criterion
in Refs.  \cite{PhysRevA.71.034306, PhysRevLett.104.070402} for $\rho=|N\>\<N|$.  Moreover it is
interesting to show that the technique applied in the derivation of Eq.~\eqref{eq:boundfordelta} can
be used to answer an open question raised in Ref.  \cite{PhysRevLett.104.070402}.
The conjecture is that, given two different composite Bosons $c_1 = \sum_{i} f_1(i) \psi_i \phi_i$ and
$c_2 = \sum_{i} f_2(i) \psi_i \phi_i$ such that $ \sum_{i} f_1(i) f_2(i)^* =0$, the commutation
relation $[c_1,c_2^\dag ]$ should vanish as the two purities $P_1$ and $P_2$ ($P_a = \sum_{i=1}^N
|f_a(i)|^4$) decrease.  Since $[c_1,c_2^\dag ] = - \sum_i f_1(i) f_2(i)^* (\psi_i^\dag \psi_i +
\phi_i^\dag \phi_i )$ we have
\begin{equation}
|\< [c_1,c_2^\dag ]\>| \leq \sum_x \sqrt{\<\Gamma^{(1)}_x \>
  \<\Gamma^{(2)}_x \> }, 
\end{equation}
by the same reasoning that we followed in the derivation of Eq.  \eqref{eq:boundfordelta}.
Combining this last inequality with the condition $ \< N |\Gamma^{(i)}_x | N\> \leq NP$ we have
$|\<N| [c_1,c_2^\dag ]|N\>| \leq 2NP $ which proves the conjecture.

\section{Phenomenological analysis}\label{sec:phen-analys}

We now investigate the new phenomenology predicted from
the modified Maxwell equations  \eqref{eq:maxwell2} and the
modified commutation relations  \eqref{eq:basiccommutation},
with a particular focus on practically testable effects. 

Let us first have a closer look at the 
dynamics described by Eq. \eqref{eq:maxwell}.
If $\bvec{u}_+$ and  $\bvec{u}_-$
are the two eigenvectors
of the matrix $\Exp [(   2 \bvec{n}_{\frac{\bk}{2}}
      \cdot \bvec{J} ) t ]$, corresponding to eigenvalues
$e^{\mp i2 |\bvec{n}_{\frac{\bk}{2}}|t} $,
Eq. \eqref{eq:maxwell} can be written as
\begin{align}
  \label{eq:maxwellsolved}
  \bvec{F}_T(\bk,t) =
 e^{-i2|\bvec{n}_{\frac{\bk}{2}}|t} \gamma_+(\bk) \bvec{u}_+
+
 e^{i2|\bvec{n}_{\frac{\bk}{2}}|t} \gamma_-(\bk) \bvec{u}_-
\end{align}
where the corresponding polarization operators
$\gamma_\pm(\bk)$ are defined according to  Eq. \eqref{eq:polarization}.
According to Eq. \eqref{eq:maxwellsolved}
the angular frequency of the electromagnetic waves 
is given by the modified dispersion relation
\begin{align}
\label{eq:modifieddisprelmax}
\omega(\bk) = 2 | \bvec{n}_{\tfrac{\bk}{2}} |   .
\end{align}
The usual relation $\omega(\bk) =  | \bk  |  $
is recovered in the $| \bk  | \ll 1$ regime.
The speed of light is the group velocity of the electromagnetic 
waves, i.e.~the gradient of the dispersion relation. The major consequence 
of Eq. \eqref{eq:modifieddisprelmax} is that the speed of light depends on 
the value of $\bk$, as for Maxwell's equations in a dispersive medium. 

The phenomenon of a $\bk$-dependent speed of light was already analyzed in the in the context of
quantum gravity where many authors considered the hypothesis that the existence of an invariant
length (the Planck scale) could manifest itself in terms of modified dispersion relations
\cite{ellis1992string,lukierski1995classical,Quantidischooft1996,amelino2001testable,PhysRevLett.88.190403}.
In these models the $\bk$-dependent speed of light $c(\bk)$, at the leading order in $k :=| \bk |$,
is expanded as $c(\bk) \approx 1 \pm \xi k^{\alpha}$, where $\xi $ is a numerical factor of order
$1$, while $\alpha$ is an integer.  This is exactly what happens in our framework, where the
intrinsic discreteness of the quantum cellular automata $A^\pm$ leads to the dispersion relation of
Eq. \eqref{eq:modifieddisprelmax} from which the following $\bk$-dependent speed of light
\begin{align} \label{eq:freqdepsol}
  c^\mp(\bk) \approx 1 \pm 3\frac{k_x k_y k_z}{|\bk|^2} \approx
 1 \pm \tfrac{1}{\sqrt{3}}k,
\end{align}
can be obtained by computing the modulus of the group velocity and power expanding in $\bk$ with the
assumption $ k_x = k_y = k_z = \tfrac{1}{\sqrt{3}} k $, ($k = |\bk|$). It is interesting to observe
that depending on the automaton $A^{+}(\bk)$ of $A^{-}(\bk)$ in Eq. \eqref{eq:weylautomata2} we
obtain corrections to the speed of light with opposite sign. Moreover the correction is not
isotropic and can be superluminal, though uniformly bounded for all $\bk$ as shown for the Weyl
automaton in Ref. \cite{d2013derivation}.

Models leading to modified dispersion relations recently received attention because they allow one
to derive falsifiable predictions of the Plank scale hypothesis. These can be experimentally tested
in the astrophysical domain, where the tiny corrections to the usual relativistic dynamics can be
magnified by the huge time of flight.  For example, observations of the arrival times of pulses
originated at cosmological distances, like in some $\gamma$-ray
bursts\cite{amelino1998tests,abdo2009limit,vasileiou2013constraints,amelino2009prospects}, are now
approaching a sufficient sensitivity to detect corrections to the relativistic dispersion relation
of the same order as in Eq. \eqref{eq:freqdepsol}.

\begin{figure}[ht]
  \begin{center}
    \includegraphics[width=8cm ]{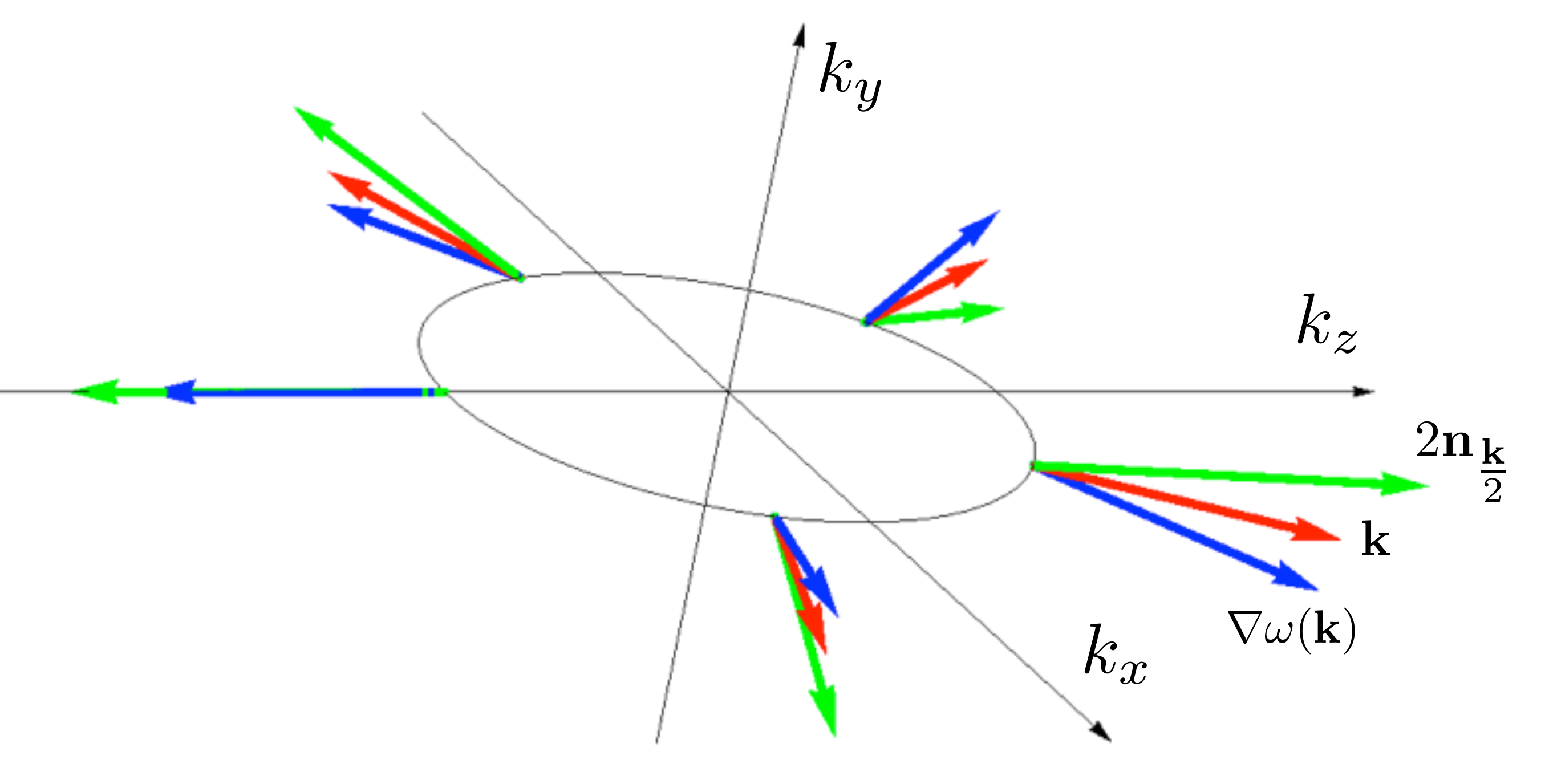}
    \caption{(colors online) The graphics shows the vector $2\bvec{n}_{\tfrac{\bk}{2}}$ (in green),
      which is orthogonal to the polarization plane, the wavevector $\bk$ (in red) and the group
      velocity $\nabla \omega (\bk)$ (in blue) as function of $\bk$ for the value $|\bk|= 0.8$ and
      different directions. Notice that the three vectors are not parallel and the angles between
      them depend on $\bk$. Such anisotropic behavior can be traced back to the anisotropy of the
      dispersion relation of the Weyl automaton.}
    \label{fig:relvectors}
  \end{center}
\end{figure}
A second distinguishing feature of Eq. \eqref{eq:maxwell2} is that the polarization plane is neither
orthogonal to the wavevector, nor to the group velocity, which means that the electromagnetic waves
are no longer exactly transverse (see Figs.~\ref{fig:elmwave} and \ref{fig:relvectors}).  However
the angle $\theta$ between the polarization plane and the plane orthogonal to $\bk$ or
$\nabla\omega(\bk)$ is of the order $\theta \approx 2k$, which gives $10^{-15}\mathrm{rad}$ for a
$\gamma$-ray wavelength, a precision which is not reachable by the present technology. Since for a
fixed $\bk$ the polarization plane is constant, exploiting greater distances and longer times does
not help in magnifying this deviation from the usual electromagnetic theory.

Finally, the third phenomenological consequence of our modelling is that, since the photon is
described as a composite Boson, deviations from the usual Bosonic statistics are in order.  As we
proved in Section \ref{sec:photons-as-composite}, the choice of the function ${f}_\bk(\bvec{q})$
determines the regime where the composite photon can be approximately treated as a Boson.  However,
independently on the details of function ${f}_\bk(\bvec{q})$ one can easily see that a Fermionic
saturation of the Boson is not visible, e.g. for the most powerful laser \cite{dunne2007high} one
has a approximately an Avogadro number of photons in $10^{-15}$cm${}^3$, whereas in the same volume
on has around $10^{90}$ Fermionic modes.


Another test for the composite nature of photons is provided by the prediction of
deviations from the Planck's distribution in Blackbody
radiation experiments. A similar analysis was carried out in
Ref. \cite{perkins2002quasibosons}, where the author showed that the predicted 
deviation from Planck's law is less than one part over $10^{-8}$, 
well beyond the sensitivity of present day experiments.

\section{Conclusions}
\label{sec:conclusions}
In this paper we derive a complete theoretical framework of the free quantum radiation field at the
Planck scale, based on a quantum Weyl automaton derived from first principles in Ref.
\cite{d2013derivation}. Differently from previous arguments based just on discreteness of geometry,
the present approach provides fully quantum theoretical treatment that allows for precise
observational predictions which involve electromagnetic radiation, e.~g. about deep-space
astrophysical sources. Within the present framework the electromagnetic field emerges from two
correlated massless Fermionic fields whose evolution is given by the Weyl automaton. Then the
electric and magnetic field are described in terms of bilinear operators of the two constituent
Fermionic fields.  This framework recalls the so-called ``neutrino theory of light'' considered in
Refs. \cite{de1934nouvelle,jordan1935neutrinotheorie,kronig1936relativistically,perkins1972statistics,perkins2002quasibosons}.

The automaton evolution leads to a set of modified Maxwell's equations whose dynamics differs from
the usual one for ultra-high wavevectors. This model predicts a longitudinal component of the
polarization and a $\bk$-dependent speed of light. This last effect could be observed by measuring
the arrival times of light originated at cosmological distances, like in some $\gamma$-ray bursts,
exploiting the huge distance scale to magnify the tiny corrective terms to the relativistic
kinematics.  This prediction agrees with the one presented in Ref. \cite{amelino1998tests} where
$\gamma$-ray bursts were for the first time considered as tests for physical models with
non-Lorentzian dispersion relations. Within this perspective, our quantum cellular automaton singles
out a specific modified dispersion relation as emergent from a Planck-scale microscopic dynamics.

Another major feature of the proposed model, is the composite nature of the photon which leads to a
modification of the Bosonic commutation relations. Because of the Fermionic structure of the photon
we expect that the Pauli exclusion principle could cause a saturation effects when a critical energy
density is achieved.  However, an order of magnitude estimation shows that the effect is very far
from being detectable with the current laser technology.

As a spin-off of the analysis of the composite nature of the photons, we proved a result that
strenghten the thesis that the amount of entanglement quantifies whether a pair of Fermions can be
treated as a Boson \cite{PhysRevA.71.034306,PhysRevLett.104.070402}.  Indeed we showed that, even in
the case of several composite Bosons, the amount of entanglement for each pair is a good measure of
how much the different pair of Fermions can be treated as independent Bosons. This question was
proposed as an open problem in Ref. \cite{PhysRevLett.104.070402}.

The results of this work leave a lot of room for future investigation. The major question is the
study of how symmetry transformations can be represented in the model.  The scenario we considered
is restricted to a fixed reference frame and in order to properly recover the standard theory we
should discuss how the Poincar\`{e} group acts on our physical model. This analysis could be done
following the lines of Ref. \cite{bibeau2013doubly} where it is shown how a QCA dynamical model is
compatible with a deformed relativity model \cite{amelino2002relativity,PhysRevLett.88.190403} which exhibits a non-linear
action of the Poincar\`{e} group.

\acknowledgements{
This work has been supported in part by the Templeton Foundation under
the project ID\# 43796 {\em A Quantum-Digital Universe}. }

\appendix

\section{Proof of Eq. \eqref{eq:approxU0}}
\label{sec:proof-eq.-eqref}
Given two vector $\bvec{a},\bvec{a}' \in \mathbb{R}^3$,
we define
\begin{align}
&U =  R_\bvec{-a}R_{\bvec{a}+\bvec{a}'}\\ 
&R_\bvec{-a} =  \exp(i \bvec{a} \cdot \boldsymbol{\sigma} \, t)
\nonumber \\
&R_{\bvec{a}+\bvec{a}'}\exp  ( -i \left(\bvec{a}+ \bvec{a}' \right) \cdot
 \boldsymbol{\sigma} \, t  ), \nonumber . 
\end{align}
By explicit computation $R_{\bvec{a}+\bvec{a}'}$ can be written as
\begin{align}
 &R_{\bvec{a}+\bvec{a}'}=\exp  ( -i (\bvec{a}+ \bvec{a}' )
 \cdot \boldsymbol{\sigma} \, t  ) = \nonumber \\
&=  \exp  (-i t |\bvec{a}+ \bvec{a}' | 
  \bvec{e}_{\bvec{a}+\bvec{a}'} \cdot \boldsymbol{\sigma}  ) = \nonumber \\
&=I \cos(|\bvec{a}+ \bvec{a}' |t) -i
\sin\left(|\left(\bvec{a}+ \bvec{a}' \right)|t\right)
\bvec{e}_{\bvec{a}+\bvec{a}'} \cdot \boldsymbol{\sigma}    \label{eq:decompoR}
\end{align}
where we introduced $\bvec{e}_\bvec{a} = \frac{\bvec{a}}{|\bvec{a}|} $ and $  \bvec{e}_{\bvec{a}+\bvec{a}'}  = \frac{\bvec{a}+\bvec{a}'}{|\bvec{a}+\bvec{a}'|}$.
For $|\bvec{a}'| \ll |\bvec{a}|$
 we have
 \begin{align}
   \begin{split}
        \label{eq:approxmod}
  & |\bvec{a} + \bvec{a}'| = \sqrt{|\bvec{a}|^2 + |\bvec{a}'|^2 + 2
     \bvec{a}\cdot \bvec{a}'} =\\
 &=|\bvec{a}| + \frac{\bvec{a}\cdot
     \bvec{a}'}{|\bvec{a}|^2} + O\left( \tfrac{ |\bvec{a}'|^2}{ |\bvec{a}|^2} \right)
   \end{split}
  \end{align}
and
\begin{align}
|\bvec{e}_{\bvec{a}+\bvec{a}'} - \bvec{e}_\bvec{a}| &= 
\left|\tfrac{\bvec{a}+\bvec{a}'}{|\bvec{a}+\bvec{a}'|} -
  \tfrac{\bvec{a}}{|\bvec{a}|} \right|=
\left|\tfrac{  
 \bvec{a} (-|\bvec{a}+\bvec{a}'|+|\bvec{a}|)  +
    \bvec{a}' |\bvec{a}|
 }
{|\bvec{a}+\bvec{a}'||\bvec{a}|} 
 \right|\leq\nonumber\\
&\leq
\tfrac{|\bvec{a}'|}{|\bvec{a} + \bvec{a}'|} +
1-   \tfrac{|\bvec{a}|}{|\bvec{a} + \bvec{a}'|} =
O\left(\tfrac{|\bvec{a}'|}{|\bvec{a} |}\right).
 \label{eq:apprxversor}
\end{align}
Then, for $|\bvec a'|\ll|\bvec a|$ we obtain
\begin{align}
&R_{\bvec{a}+\bvec{a}'}  =
 I \cos\left(\left( |\bvec{a}| + \tfrac{\bvec{a}\cdot
       \bvec{a}'}{|\bvec{a}|}\right)t\right) + \nonumber
\\  &-i \sin\left(\left( |\bvec{a}| + \tfrac{\bvec{a}\cdot \bvec{a}'}{|\bvec{a}|}\right)t\right)  \bvec{e}_{\bvec{a}} \cdot \boldsymbol{\sigma} +
\Lambda'(\bvec{a},\bvec{a}') + \Theta'(\bvec{a},\bvec{a}')= \nonumber \\
& \exp\left(-i t \left( |\bvec{a}| + \tfrac{\bvec{a}\cdot
      \bvec{a}'}{|\bvec{a}|}\right) \bvec{e}_\bvec{a} \cdot
  \boldsymbol{\sigma}\right) + 
\Lambda'(\bvec{a},\bvec{a}') + \Theta'(\bvec{a},\bvec{a}',t) \nonumber
\end{align}
where $\Lambda'(\bvec{a},\bvec{a}')$ + 
$\Theta'(\bvec{a},\bvec{a}',t)$
are a couple of operators such that
\begin{align}
|\Lambda'(\bvec{a},\bvec{a}')| =
O\left(\tfrac{|\bvec{a}'|}{|\bvec{a}|}\right), \quad 
|\Theta'(\bvec{a},\bvec{a}')| =
O\left(\tfrac{|\bvec{a}'|^2}{|\bvec{a}|^2}t\right) 
\nonumber
\end{align}
from which we finally get
\begin{align}
  \label{eq:approximU}
& U =   \exp\left(-it \frac{\bvec{a}\cdot
    \bvec{a}'}{|\bvec{a}|}\bvec{e}_\bvec{a} \cdot \boldsymbol{\sigma}\right)
 + 
\Lambda(\bvec{a},\bvec{a}') + \Theta(\bvec{a},\bvec{a}',t) \nonumber
\\
&|\Lambda(\bvec{a},\bvec{a}')| =
O\left(\tfrac{|\bvec{a}'|}{|\bvec{a}|}\right), \quad 
|\Theta(\bvec{a},\bvec{a}')| =
O\left(\tfrac{|\bvec{a}'|^2}{|\bvec{a}|^2}t\right) 
\end{align}
which leads to Eq. \eqref{eq:approxU0} if we identify $\bvec{a}=\bvec{n}_\bk$, 
$\bvec{a}'=\bvec{l}_{\bk, \bq}$.

\section{Proof of Eq. \eqref{eq:transverse}}
\label{sec:proof-eq.-eqref-1}
Let us introduce the vectors $\bvec u^1_\bk, \bvec u^2_\bk \in \mathbb{R}^3$
such that 
\begin{align}
  \label{eq:vectors}
  \begin{split}
    \bvec u^1_\bk \cdot \bvec n_\bk=0 \quad
\bvec u^2_\bk :=\bvec{e}_\bk \times\bvec u^1_\bk \quad
\bvec{e}_\bk := {|\bvec{n}_{\tfrac{\bk}{2}}|}^{-1}{\bvec{n}_{\tfrac{\bk}{2}}}.
  \end{split}
\end{align}
The transverse field $ \tilde{\bvec{F}}_T(\bk,t)$
defined in Eq. \eqref{eq:transverse} can then be written in the basis $\{\bvec u^i_\bk\}$ as
\begin{align}
  \begin{split}
      \tilde{\bvec{F}}_T(\bk,t) =  
      \begin{pmatrix}
        \bvec u^1_{\tfrac{\bk}2}\cdot\tilde{\bvec F}^{\bvec{u}_1}(\bk,t) \\
     \bvec u^2_{\tfrac{\bk}2}\cdot\tilde{\bvec F}^{\bvec{u}_2}(\bk,t) 
      \end{pmatrix}
  \end{split}
\end{align}
Reminding the definition \eqref{eq:ftilde} we have
\begin{align}
  &\bvec u^i_{\tfrac{\bk}2}\cdot\tilde{\bvec F}(\bk,t) = 
\int\!\!\frac{\d\bq}{(2\pi)^3} f_{\bk}(\bq)
{{\varphi}}^T 
\left(\tfrac{\bk}{2}-\bq\right)
 Q^i
{\psi}
 \left(\tfrac{\bk}{2}+\bq\right) \nonumber\\
&\qquad Q^i (\bk,\bq,t):=  ({U}^{\bk,t}_{\tfrac{\bk}2-\bq})^\dag
\bvec u^i_{\tfrac{\bk}2} \cdot\boldsymbol{\sigma}
  {U}^{\bk,t}_{\tfrac{\bk}2+\bq}. \label{eq:operatO}
\end{align}
If we insert Eq. \eqref{eq:approxU0}, which can be written as
\begin{align}
&   {U}^{\bk,t}_{\tfrac{\bk}2\pm\bq} = R_{\pm\xi \bvec{e}} + O\big(  \tfrac{\bar{q}(\bk)}{|\bvec{n}_{\frac{\bk}{2}}|}
\big) \\
&R_{\pm\xi \bvec{e}} := \exp( \pm i \xi \bvec{e} \cdot
   \boldsymbol{\sigma}) \quad
\xi := c_{\bk,\bq} t\;,
\end{align}
inside Eq. \eqref{eq:operatO} we have
\begin{align}
 & Q^i (\bk,\bq,t)=
  R_{-\xi \bvec{e}}
\bvec u^i_{\tfrac{\bk}2} \cdot\boldsymbol{\sigma}
  R_{-\xi \bvec{e}}
 + O\big(
   \tfrac{\bar{q}(\bk)}{|\bvec{n}_{\frac{\bk}{2}}|}\big)= \nonumber\\  
&=\bvec u^i_{\tfrac{\bk}2} \cdot\boldsymbol{\sigma} + O\big(
\tfrac{\bar{q}(\bk)}{|\bvec{n}_{\frac{\bk}{2}}|} \big) =
Q^i (\bk,\bq,0) + O\big(
\tfrac{\bar{q}(\bk)}{|\bvec{n}_{\frac{\bk}{2}}|}\big) 
\label{eq:operatorOev},
\end{align}
where we used the identity
\begin{align}
(\bvec{a}\cdot \boldsymbol{\sigma})
 (\bvec{b} \cdot \boldsymbol{\sigma})(\bvec{a}\cdot \boldsymbol{\sigma})=-\bvec{b} \cdot \boldsymbol{\sigma}
\end{align}
holding for $\bvec a\cdot\bvec b=0$, $|\bvec a|=|\bvec b|=1$, which implies
\begin{align}
 \exp(i \xi \bvec{e} \cdot \boldsymbol{\sigma}) 
\bvec u^i_{\tfrac{\bk}2}\cdot \boldsymbol{\sigma}
 \exp(i \xi \bvec{e} \cdot \boldsymbol{\sigma}) =
\bvec u^i_{\tfrac{\bk}2}\cdot \boldsymbol{\sigma} \quad  \forall \xi \in \mathbb{R}.
\end{align}
Inserting Eq. \eqref{eq:operatorOev} in Eq. \eqref{eq:operatO}
we
have
\begin{align}
  \label{eq:ftildeapproxevolut}
    \bvec u^i_{\tfrac{\bk}2}\cdot\tilde{\bvec F}(\bk,t) = 
   \bvec u^i_{\tfrac{\bk}2}\cdot\tilde{\bvec F}(\bk,0) +  O\big(
\tfrac{\bar{q}(\bk)}{|\bvec{n}_{\frac{\bk}{2}}|} \big) 
\quad i =1,2 \nonumber
\end{align}
which then implies
\begin{align}
\tilde{\bvec{F}}_T(\bk,t) =  \tilde{\bvec{F}}_T(\bk,0) +  O\big(
\tfrac{\bar{q}(\bk)}{|\bvec{n}_{\frac{\bk}{2}}|} \big)  ={\bvec{F}}_T(\bk)  +  O\big(
\tfrac{\bar{q}(\bk)}{|\bvec{n}_{\frac{\bk}{2}}|} \big)    \nonumber
\end{align}

\bibliography{bibliography}

\begin{thebibliography}{52}
\expandafter\ifx\csname natexlab\endcsname\relax\def\natexlab#1{#1}\fi
\expandafter\ifx\csname bibnamefont\endcsname\relax
  \def\bibnamefont#1{#1}\fi
\expandafter\ifx\csname bibfnamefont\endcsname\relax
  \def\bibfnamefont#1{#1}\fi
\expandafter\ifx\csname citenamefont\endcsname\relax
  \def\citenamefont#1{#1}\fi
\expandafter\ifx\csname url\endcsname\relax
  \def\url#1{\texttt{#1}}\fi
\expandafter\ifx\csname urlprefix\endcsname\relax\def\urlprefix{URL }\fi
\providecommand{\bibinfo}[2]{#2}
\providecommand{\eprint}[2][]{\url{#2}}

\bibitem[{\citenamefont{D'Ariano and Perinotti}(2013)}]{d2013derivation}
\bibinfo{author}{\bibfnamefont{G.~M.} \bibnamefont{D'Ariano}} \bibnamefont{and}
  \bibinfo{author}{\bibfnamefont{P.}~\bibnamefont{Perinotti}},
  \bibinfo{journal}{arXiv preprint arXiv:1306.1934}  (\bibinfo{year}{2013}).

\bibitem[{\citenamefont{von Neumann}(1966)}]{neumann1966theory}
\bibinfo{author}{\bibfnamefont{J.}~\bibnamefont{von Neumann}},
  \emph{\bibinfo{title}{Theory of self-reproducing automata}}
  (\bibinfo{publisher}{University of Illinois Press}, \bibinfo{address}{Urbana
  and London}, \bibinfo{year}{1966}).

\bibitem[{\citenamefont{Feynman}(1982)}]{feynman1982simulating}
\bibinfo{author}{\bibfnamefont{R.}~\bibnamefont{Feynman}},
  \bibinfo{journal}{International journal of theoretical physics}
  \textbf{\bibinfo{volume}{21}}, \bibinfo{pages}{467} (\bibinfo{year}{1982}).

\bibitem[{\citenamefont{Schumacher and
  Werner}(2004)}]{schumacher2004reversible}
\bibinfo{author}{\bibfnamefont{B.}~\bibnamefont{Schumacher}} \bibnamefont{and}
  \bibinfo{author}{\bibfnamefont{R.}~\bibnamefont{Werner}},
  \bibinfo{journal}{Arxiv preprint quant-ph/0405174}  (\bibinfo{year}{2004}).

\bibitem[{\citenamefont{Arrighi et~al.}(2011)\citenamefont{Arrighi, Nesme, and
  Werner}}]{arrighi2011unitarity}
\bibinfo{author}{\bibfnamefont{P.}~\bibnamefont{Arrighi}},
  \bibinfo{author}{\bibfnamefont{V.}~\bibnamefont{Nesme}}, \bibnamefont{and}
  \bibinfo{author}{\bibfnamefont{R.}~\bibnamefont{Werner}},
  \bibinfo{journal}{Journal of Computer and System Sciences}
  \textbf{\bibinfo{volume}{77}}, \bibinfo{pages}{372} (\bibinfo{year}{2011}).

\bibitem[{\citenamefont{Gross et~al.}(2012)\citenamefont{Gross, Nesme, Vogts,
  and Werner}}]{gross2012index}
\bibinfo{author}{\bibfnamefont{D.}~\bibnamefont{Gross}},
  \bibinfo{author}{\bibfnamefont{V.}~\bibnamefont{Nesme}},
  \bibinfo{author}{\bibfnamefont{H.}~\bibnamefont{Vogts}}, \bibnamefont{and}
  \bibinfo{author}{\bibfnamefont{R.}~\bibnamefont{Werner}},
  \bibinfo{journal}{Communications in Mathematical Physics} pp.
  \bibinfo{pages}{1--36} (\bibinfo{year}{2012}).

\bibitem[{\citenamefont{Grossing and Zeilinger}(1988)}]{grossing1988quantum}
\bibinfo{author}{\bibfnamefont{G.}~\bibnamefont{Grossing}} \bibnamefont{and}
  \bibinfo{author}{\bibfnamefont{A.}~\bibnamefont{Zeilinger}},
  \bibinfo{journal}{Complex Systems} \textbf{\bibinfo{volume}{2}},
  \bibinfo{pages}{197} (\bibinfo{year}{1988}).

\bibitem[{\citenamefont{Succi and Benzi}(1993)}]{succi1993lattice}
\bibinfo{author}{\bibfnamefont{S.}~\bibnamefont{Succi}} \bibnamefont{and}
  \bibinfo{author}{\bibfnamefont{R.}~\bibnamefont{Benzi}},
  \bibinfo{journal}{Physica D: Nonlinear Phenomena}
  \textbf{\bibinfo{volume}{69}}, \bibinfo{pages}{327} (\bibinfo{year}{1993}).

\bibitem[{\citenamefont{Meyer}(1996)}]{meyer1996quantum}
\bibinfo{author}{\bibfnamefont{D.}~\bibnamefont{Meyer}},
  \bibinfo{journal}{Journal of Statistical Physics}
  \textbf{\bibinfo{volume}{85}}, \bibinfo{pages}{551} (\bibinfo{year}{1996}).

\bibitem[{\citenamefont{Bialynicki-Birula}(1994)}]{bialynicki1994weyl}
\bibinfo{author}{\bibfnamefont{I.}~\bibnamefont{Bialynicki-Birula}},
  \bibinfo{journal}{Physical Review D} \textbf{\bibinfo{volume}{49}},
  \bibinfo{pages}{6920} (\bibinfo{year}{1994}).

\bibitem[{\citenamefont{Ambainis et~al.}(2001)\citenamefont{Ambainis, Bach,
  Nayak, Vishwanath, and Watrous}}]{ambainis2001one}
\bibinfo{author}{\bibfnamefont{A.}~\bibnamefont{Ambainis}},
  \bibinfo{author}{\bibfnamefont{E.}~\bibnamefont{Bach}},
  \bibinfo{author}{\bibfnamefont{A.}~\bibnamefont{Nayak}},
  \bibinfo{author}{\bibfnamefont{A.}~\bibnamefont{Vishwanath}},
  \bibnamefont{and} \bibinfo{author}{\bibfnamefont{J.}~\bibnamefont{Watrous}},
  in \emph{\bibinfo{booktitle}{Proceedings of the thirty-third annual ACM
  symposium on Theory of computing}} (\bibinfo{organization}{ACM},
  \bibinfo{year}{2001}), pp. \bibinfo{pages}{37--49}.

\bibitem[{\citenamefont{Cirac and Zoller}(2000)}]{cirac2000scalable}
\bibinfo{author}{\bibfnamefont{J.~I.} \bibnamefont{Cirac}} \bibnamefont{and}
  \bibinfo{author}{\bibfnamefont{P.}~\bibnamefont{Zoller}},
  \bibinfo{journal}{Nature} \textbf{\bibinfo{volume}{404}},
  \bibinfo{pages}{579} (\bibinfo{year}{2000}).

\bibitem[{\citenamefont{Bloch}(2004)}]{bloch2004quantum}
\bibinfo{author}{\bibfnamefont{I.}~\bibnamefont{Bloch}},
  \bibinfo{journal}{Physics World} \textbf{\bibinfo{volume}{17}},
  \bibinfo{pages}{25} (\bibinfo{year}{2004}).

\bibitem[{\citenamefont{Childs et~al.}(2003)\citenamefont{Childs, Cleve,
  Deotto, Farhi, Gutmann, and Spielman}}]{childs2003exponential}
\bibinfo{author}{\bibfnamefont{A.~M.} \bibnamefont{Childs}},
  \bibinfo{author}{\bibfnamefont{R.}~\bibnamefont{Cleve}},
  \bibinfo{author}{\bibfnamefont{E.}~\bibnamefont{Deotto}},
  \bibinfo{author}{\bibfnamefont{E.}~\bibnamefont{Farhi}},
  \bibinfo{author}{\bibfnamefont{S.}~\bibnamefont{Gutmann}}, \bibnamefont{and}
  \bibinfo{author}{\bibfnamefont{D.~A.} \bibnamefont{Spielman}}, in
  \emph{\bibinfo{booktitle}{Proceedings of the thirty-fifth annual ACM
  symposium on Theory of computing}} (\bibinfo{organization}{ACM},
  \bibinfo{year}{2003}), pp. \bibinfo{pages}{59--68}.

\bibitem[{\citenamefont{Farhi et~al.}(2007)\citenamefont{Farhi, Goldstone, and
  Gutmann}}]{farhi2007quantum}
\bibinfo{author}{\bibfnamefont{E.}~\bibnamefont{Farhi}},
  \bibinfo{author}{\bibfnamefont{J.}~\bibnamefont{Goldstone}},
  \bibnamefont{and} \bibinfo{author}{\bibfnamefont{S.}~\bibnamefont{Gutmann}},
  \bibinfo{journal}{arXiv preprint quant-ph/0702144}  (\bibinfo{year}{2007}).

\bibitem[{\citenamefont{D'Ariano}(2011)}]{darianopla}
\bibinfo{author}{\bibfnamefont{G.~M.} \bibnamefont{D'Ariano}},
  \bibinfo{journal}{Phys. Lett. A} \textbf{\bibinfo{volume}{376}}
  (\bibinfo{year}{2011}).

\bibitem[{\citenamefont{Bisio et~al.}(2012)\citenamefont{Bisio, D'Ariano, and
  Tosini}}]{BDTqcaI}
\bibinfo{author}{\bibfnamefont{A.}~\bibnamefont{Bisio}},
  \bibinfo{author}{\bibfnamefont{G.}~\bibnamefont{D'Ariano}}, \bibnamefont{and}
  \bibinfo{author}{\bibfnamefont{A.}~\bibnamefont{Tosini}},
  \bibinfo{journal}{arXiv preprint arXiv:1212.2839}  (\bibinfo{year}{2012}).

\bibitem[{\citenamefont{Farrelly and Short}(2014)}]{farrelly2014causal}
\bibinfo{author}{\bibfnamefont{T.~C.} \bibnamefont{Farrelly}} \bibnamefont{and}
  \bibinfo{author}{\bibfnamefont{A.~J.} \bibnamefont{Short}},
  \bibinfo{journal}{Physical Review A} \textbf{\bibinfo{volume}{89}},
  \bibinfo{pages}{012302} (\bibinfo{year}{2014}).

\bibitem[{\citenamefont{Arrighi et~al.}(2013)\citenamefont{Arrighi, Forets, and
  Nesme}}]{arrighi2013dirac}
\bibinfo{author}{\bibfnamefont{P.}~\bibnamefont{Arrighi}},
  \bibinfo{author}{\bibfnamefont{M.}~\bibnamefont{Forets}}, \bibnamefont{and}
  \bibinfo{author}{\bibfnamefont{V.}~\bibnamefont{Nesme}},
  \bibinfo{journal}{arXiv preprint arXiv:1307.3524}  (\bibinfo{year}{2013}).

\bibitem[{\citenamefont{'t~Hooft}(1990)}]{t1990quantization}
\bibinfo{author}{\bibfnamefont{G.}~\bibnamefont{'t~Hooft}},
  \bibinfo{journal}{Nuclear Physics B} \textbf{\bibinfo{volume}{342}},
  \bibinfo{pages}{471} (\bibinfo{year}{1990}).

\bibitem[{\citenamefont{Strauch}(2006)}]{PhysRevA.73.054302}
\bibinfo{author}{\bibfnamefont{F.~W.} \bibnamefont{Strauch}},
  \bibinfo{journal}{Phys. Rev. A} \textbf{\bibinfo{volume}{73}},
  \bibinfo{pages}{054302} (\bibinfo{year}{2006}).

\bibitem[{\citenamefont{Yepez}(2006)}]{Yepez:2006p4406}
\bibinfo{author}{\bibfnamefont{J.}~\bibnamefont{Yepez}},
  \bibinfo{journal}{Quantum Information Processing}
  \textbf{\bibinfo{volume}{4}}, \bibinfo{pages}{471} (\bibinfo{year}{2006}).

\bibitem[{\citenamefont{Bekenstein}(1973)}]{bekenstein1973black}
\bibinfo{author}{\bibfnamefont{J.~D.} \bibnamefont{Bekenstein}},
  \bibinfo{journal}{Physical Review D} \textbf{\bibinfo{volume}{7}},
  \bibinfo{pages}{2333} (\bibinfo{year}{1973}).

\bibitem[{\citenamefont{Hawking}(1975)}]{hawking1975particle}
\bibinfo{author}{\bibfnamefont{S.~W.} \bibnamefont{Hawking}},
  \bibinfo{journal}{Communications in mathematical physics}
  \textbf{\bibinfo{volume}{43}}, \bibinfo{pages}{199} (\bibinfo{year}{1975}).

\bibitem[{\citenamefont{Ellis et~al.}(1992)\citenamefont{Ellis, Mavromatos, and
  Nanopoulos}}]{ellis1992string}
\bibinfo{author}{\bibfnamefont{J.}~\bibnamefont{Ellis}},
  \bibinfo{author}{\bibfnamefont{N.}~\bibnamefont{Mavromatos}},
  \bibnamefont{and} \bibinfo{author}{\bibfnamefont{D.~V.}
  \bibnamefont{Nanopoulos}}, \bibinfo{journal}{Physics Letters B}
  \textbf{\bibinfo{volume}{293}}, \bibinfo{pages}{37} (\bibinfo{year}{1992}).

\bibitem[{\citenamefont{Lukierski et~al.}(1995)\citenamefont{Lukierski, Ruegg,
  and Zakrzewski}}]{lukierski1995classical}
\bibinfo{author}{\bibfnamefont{J.}~\bibnamefont{Lukierski}},
  \bibinfo{author}{\bibfnamefont{H.}~\bibnamefont{Ruegg}}, \bibnamefont{and}
  \bibinfo{author}{\bibfnamefont{W.~J.} \bibnamefont{Zakrzewski}},
  \bibinfo{journal}{Annals of Physics} \textbf{\bibinfo{volume}{243}},
  \bibinfo{pages}{90} (\bibinfo{year}{1995}).

\bibitem[{\citenamefont{'t~Hooft}(1996)}]{Quantidischooft1996}
\bibinfo{author}{\bibfnamefont{G.}~\bibnamefont{'t~Hooft}},
  \bibinfo{journal}{Class. Quantum Grav.} \textbf{\bibinfo{volume}{13}},
  \bibinfo{pages}{1023} (\bibinfo{year}{1996}).

\bibitem[{\citenamefont{Amelino-Camelia
  et~al.}(1998)\citenamefont{Amelino-Camelia, Ellis, Mavromatos, Nanopoulos,
  and Sarkar}}]{amelino1998tests}
\bibinfo{author}{\bibfnamefont{G.}~\bibnamefont{Amelino-Camelia}},
  \bibinfo{author}{\bibfnamefont{J.}~\bibnamefont{Ellis}},
  \bibinfo{author}{\bibfnamefont{N.}~\bibnamefont{Mavromatos}},
  \bibinfo{author}{\bibfnamefont{D.~V.} \bibnamefont{Nanopoulos}},
  \bibnamefont{and} \bibinfo{author}{\bibfnamefont{S.}~\bibnamefont{Sarkar}},
  \bibinfo{journal}{Nature} \textbf{\bibinfo{volume}{393}},
  \bibinfo{pages}{763} (\bibinfo{year}{1998}).

\bibitem[{\citenamefont{Amelino-Camelia}(2001)}]{amelino2001testable}
\bibinfo{author}{\bibfnamefont{G.}~\bibnamefont{Amelino-Camelia}},
  \bibinfo{journal}{Physics Letters B} \textbf{\bibinfo{volume}{510}},
  \bibinfo{pages}{255} (\bibinfo{year}{2001}).

\bibitem[{\citenamefont{Amelino-Camelia and Piran}(2001)}]{amelino2001planck}
\bibinfo{author}{\bibfnamefont{G.}~\bibnamefont{Amelino-Camelia}}
  \bibnamefont{and} \bibinfo{author}{\bibfnamefont{T.}~\bibnamefont{Piran}},
  \bibinfo{journal}{Physical Review D} \textbf{\bibinfo{volume}{64}},
  \bibinfo{pages}{036005} (\bibinfo{year}{2001}).

\bibitem[{\citenamefont{Magueijo and Smolin}(2002)}]{PhysRevLett.88.190403}
\bibinfo{author}{\bibfnamefont{J.}~\bibnamefont{Magueijo}} \bibnamefont{and}
  \bibinfo{author}{\bibfnamefont{L.}~\bibnamefont{Smolin}},
  \bibinfo{journal}{Phys. Rev. Lett.} \textbf{\bibinfo{volume}{88}},
  \bibinfo{pages}{190403} (\bibinfo{year}{2002}).

\bibitem[{\citenamefont{Christiansen et~al.}(2006)\citenamefont{Christiansen,
  Ng, and van Dam}}]{PhysRevLett.96.051301}
\bibinfo{author}{\bibfnamefont{W.~A.} \bibnamefont{Christiansen}},
  \bibinfo{author}{\bibfnamefont{Y.~J.} \bibnamefont{Ng}}, \bibnamefont{and}
  \bibinfo{author}{\bibfnamefont{H.}~\bibnamefont{van Dam}},
  \bibinfo{journal}{Phys. Rev. Lett.} \textbf{\bibinfo{volume}{96}},
  \bibinfo{pages}{051301} (\bibinfo{year}{2006}),
  \urlprefix\url{http://link.aps.org/doi/10.1103/PhysRevLett.96.051301}.

\bibitem[{\citenamefont{Ellis and Mavromatos}(2013)}]{ellis2013probes}
\bibinfo{author}{\bibfnamefont{J.}~\bibnamefont{Ellis}} \bibnamefont{and}
  \bibinfo{author}{\bibfnamefont{N.~E.} \bibnamefont{Mavromatos}},
  \bibinfo{journal}{Astroparticle Physics} \textbf{\bibinfo{volume}{43}},
  \bibinfo{pages}{50} (\bibinfo{year}{2013}).

\bibitem[{\citenamefont{De~Broglie}(1934)}]{de1934nouvelle}
\bibinfo{author}{\bibfnamefont{L.}~\bibnamefont{De~Broglie}},
  \emph{\bibinfo{title}{Une nouvelle conception de la lumi{\`e}re}}, vol.
  \bibinfo{volume}{181} (\bibinfo{publisher}{Hermamm \& Cie},
  \bibinfo{year}{1934}).

\bibitem[{\citenamefont{Jordan}(1935)}]{jordan1935neutrinotheorie}
\bibinfo{author}{\bibfnamefont{P.}~\bibnamefont{Jordan}},
  \bibinfo{journal}{Zeitschrift f{\"u}r Physik} \textbf{\bibinfo{volume}{93}},
  \bibinfo{pages}{464} (\bibinfo{year}{1935}).

\bibitem[{\citenamefont{Kronig}(1936)}]{kronig1936relativistically}
\bibinfo{author}{\bibfnamefont{R.~d.~L.} \bibnamefont{Kronig}},
  \bibinfo{journal}{Physica} \textbf{\bibinfo{volume}{3}},
  \bibinfo{pages}{1120} (\bibinfo{year}{1936}).

\bibitem[{\citenamefont{Perkins}(1972)}]{perkins1972statistics}
\bibinfo{author}{\bibfnamefont{W.}~\bibnamefont{Perkins}},
  \bibinfo{journal}{Physical Review D} \textbf{\bibinfo{volume}{5}},
  \bibinfo{pages}{1375} (\bibinfo{year}{1972}).

\bibitem[{\citenamefont{Perkins}(2002)}]{perkins2002quasibosons}
\bibinfo{author}{\bibfnamefont{W.}~\bibnamefont{Perkins}},
  \bibinfo{journal}{International Journal of Theoretical Physics}
  \textbf{\bibinfo{volume}{41}}, \bibinfo{pages}{823} (\bibinfo{year}{2002}).

\bibitem[{\citenamefont{Pryce}(1938)}]{pryce1938neutrino}
\bibinfo{author}{\bibfnamefont{M.}~\bibnamefont{Pryce}},
  \bibinfo{journal}{Proceedings of the Royal Society of London. Series A.
  Mathematical and Physical Sciences} \textbf{\bibinfo{volume}{165}},
  \bibinfo{pages}{247} (\bibinfo{year}{1938}).

\bibitem[{\citenamefont{Chudzicki et~al.}(2010)\citenamefont{Chudzicki, Oke,
  and Wootters}}]{PhysRevLett.104.070402}
\bibinfo{author}{\bibfnamefont{C.}~\bibnamefont{Chudzicki}},
  \bibinfo{author}{\bibfnamefont{O.}~\bibnamefont{Oke}}, \bibnamefont{and}
  \bibinfo{author}{\bibfnamefont{W.~K.} \bibnamefont{Wootters}},
  \bibinfo{journal}{Phys. Rev. Lett.} \textbf{\bibinfo{volume}{104}},
  \bibinfo{pages}{070402} (\bibinfo{year}{2010}),
  \urlprefix\url{http://link.aps.org/doi/10.1103/PhysRevLett.104.070402}.

\bibitem[{\citenamefont{de~Cornulier et~al.}(2007)\citenamefont{de~Cornulier,
  Tessera, and Valette}}]{Cornulier07}
\bibinfo{author}{\bibfnamefont{Y.}~\bibnamefont{de~Cornulier}},
  \bibinfo{author}{\bibfnamefont{R.}~\bibnamefont{Tessera}}, \bibnamefont{and}
  \bibinfo{author}{\bibfnamefont{A.}~\bibnamefont{Valette}},
  \bibinfo{journal}{GAFA Geometric And Functional Analysis}
  \textbf{\bibinfo{volume}{17}}, \bibinfo{pages}{770} (\bibinfo{year}{2007}),
  ISSN \bibinfo{issn}{1016-443X},
  \urlprefix\url{http://dx.doi.org/10.1007/s00039-007-0604-0}.

\bibitem[{\citenamefont{Combescot and Tanguy}(2001)}]{combescot2001new}
\bibinfo{author}{\bibfnamefont{M.}~\bibnamefont{Combescot}} \bibnamefont{and}
  \bibinfo{author}{\bibfnamefont{C.}~\bibnamefont{Tanguy}},
  \bibinfo{journal}{EPL (Europhysics Letters)} \textbf{\bibinfo{volume}{55}},
  \bibinfo{pages}{390} (\bibinfo{year}{2001}).

\bibitem[{\citenamefont{Rombouts et~al.}(2002)\citenamefont{Rombouts, Van~Neck,
  Peirs, and Pollet}}]{rombouts2002maximum}
\bibinfo{author}{\bibfnamefont{S.}~\bibnamefont{Rombouts}},
  \bibinfo{author}{\bibfnamefont{D.}~\bibnamefont{Van~Neck}},
  \bibinfo{author}{\bibfnamefont{K.}~\bibnamefont{Peirs}}, \bibnamefont{and}
  \bibinfo{author}{\bibfnamefont{L.}~\bibnamefont{Pollet}},
  \bibinfo{journal}{Modern Physics Letters A} \textbf{\bibinfo{volume}{17}},
  \bibinfo{pages}{1899} (\bibinfo{year}{2002}).

\bibitem[{\citenamefont{Avancini et~al.}(2003)\citenamefont{Avancini,
  Marinelli, and Krein}}]{avancini2003compositeness}
\bibinfo{author}{\bibfnamefont{S.}~\bibnamefont{Avancini}},
  \bibinfo{author}{\bibfnamefont{J.}~\bibnamefont{Marinelli}},
  \bibnamefont{and} \bibinfo{author}{\bibfnamefont{G.}~\bibnamefont{Krein}},
  \bibinfo{journal}{Journal of Physics A: Mathematical and General}
  \textbf{\bibinfo{volume}{36}}, \bibinfo{pages}{9045} (\bibinfo{year}{2003}).

\bibitem[{\citenamefont{Combescot et~al.}(2003)\citenamefont{Combescot,
  Leyronas, and Tanguy}}]{combescot2003n}
\bibinfo{author}{\bibfnamefont{M.}~\bibnamefont{Combescot}},
  \bibinfo{author}{\bibfnamefont{X.}~\bibnamefont{Leyronas}}, \bibnamefont{and}
  \bibinfo{author}{\bibfnamefont{C.}~\bibnamefont{Tanguy}},
  \bibinfo{journal}{The European Physical Journal B-Condensed Matter and
  Complex Systems} \textbf{\bibinfo{volume}{31}}, \bibinfo{pages}{17}
  (\bibinfo{year}{2003}).

\bibitem[{\citenamefont{Law}(2005)}]{PhysRevA.71.034306}
\bibinfo{author}{\bibfnamefont{C.~K.} \bibnamefont{Law}},
  \bibinfo{journal}{Phys. Rev. A} \textbf{\bibinfo{volume}{71}},
  \bibinfo{pages}{034306} (\bibinfo{year}{2005}),
  \urlprefix\url{http://link.aps.org/doi/10.1103/PhysRevA.71.034306}.

\bibitem[{\citenamefont{Abdo et~al.}(2009)\citenamefont{Abdo, Ackermann,
  Ajello, Asano, Atwood, Axelsson, Baldini, Ballet, Barbiellini, Baring
  et~al.}}]{abdo2009limit}
\bibinfo{author}{\bibfnamefont{A.}~\bibnamefont{Abdo}},
  \bibinfo{author}{\bibfnamefont{M.}~\bibnamefont{Ackermann}},
  \bibinfo{author}{\bibfnamefont{M.}~\bibnamefont{Ajello}},
  \bibinfo{author}{\bibfnamefont{K.}~\bibnamefont{Asano}},
  \bibinfo{author}{\bibfnamefont{W.}~\bibnamefont{Atwood}},
  \bibinfo{author}{\bibfnamefont{M.}~\bibnamefont{Axelsson}},
  \bibinfo{author}{\bibfnamefont{L.}~\bibnamefont{Baldini}},
  \bibinfo{author}{\bibfnamefont{J.}~\bibnamefont{Ballet}},
  \bibinfo{author}{\bibfnamefont{G.}~\bibnamefont{Barbiellini}},
  \bibinfo{author}{\bibfnamefont{M.}~\bibnamefont{Baring}},
  \bibnamefont{et~al.}, \bibinfo{journal}{Nature}
  \textbf{\bibinfo{volume}{462}}, \bibinfo{pages}{331} (\bibinfo{year}{2009}).

\bibitem[{\citenamefont{Vasileiou et~al.}(2013)\citenamefont{Vasileiou,
  Jacholkowska, Piron, Bolmont, Couturier, Granot, Stecker, Cohen-Tanugi, and
  Longo}}]{vasileiou2013constraints}
\bibinfo{author}{\bibfnamefont{V.}~\bibnamefont{Vasileiou}},
  \bibinfo{author}{\bibfnamefont{A.}~\bibnamefont{Jacholkowska}},
  \bibinfo{author}{\bibfnamefont{F.}~\bibnamefont{Piron}},
  \bibinfo{author}{\bibfnamefont{J.}~\bibnamefont{Bolmont}},
  \bibinfo{author}{\bibfnamefont{C.}~\bibnamefont{Couturier}},
  \bibinfo{author}{\bibfnamefont{J.}~\bibnamefont{Granot}},
  \bibinfo{author}{\bibfnamefont{F.}~\bibnamefont{Stecker}},
  \bibinfo{author}{\bibfnamefont{J.}~\bibnamefont{Cohen-Tanugi}},
  \bibnamefont{and} \bibinfo{author}{\bibfnamefont{F.}~\bibnamefont{Longo}},
  \bibinfo{journal}{Physical Review D} \textbf{\bibinfo{volume}{87}},
  \bibinfo{pages}{122001} (\bibinfo{year}{2013}).

\bibitem[{\citenamefont{Amelino-Camelia and
  Smolin}(2009)}]{amelino2009prospects}
\bibinfo{author}{\bibfnamefont{G.}~\bibnamefont{Amelino-Camelia}}
  \bibnamefont{and} \bibinfo{author}{\bibfnamefont{L.}~\bibnamefont{Smolin}},
  \bibinfo{journal}{Physical Review D} \textbf{\bibinfo{volume}{80}},
  \bibinfo{pages}{084017} (\bibinfo{year}{2009}).

\bibitem[{\citenamefont{Dunne}(2007)}]{dunne2007high}
\bibinfo{author}{\bibfnamefont{M.}~\bibnamefont{Dunne}}, in
  \emph{\bibinfo{booktitle}{Conference on Lasers and Electro-Optics/Pacific
  Rim}} (\bibinfo{organization}{Optical Society of America},
  \bibinfo{year}{2007}), pp. \bibinfo{pages}{1--2}.

\bibitem[{\citenamefont{Bibeau-Delisle
  et~al.}(2013)\citenamefont{Bibeau-Delisle, Bisio, D'Ariano, Perinotti, and
  Tosini}}]{bibeau2013doubly}
\bibinfo{author}{\bibfnamefont{A.}~\bibnamefont{Bibeau-Delisle}},
  \bibinfo{author}{\bibfnamefont{A.}~\bibnamefont{Bisio}},
  \bibinfo{author}{\bibfnamefont{G.~M.} \bibnamefont{D'Ariano}},
  \bibinfo{author}{\bibfnamefont{P.}~\bibnamefont{Perinotti}},
  \bibnamefont{and} \bibinfo{author}{\bibfnamefont{A.}~\bibnamefont{Tosini}},
  \bibinfo{journal}{arXiv preprint arXiv:1310.6760}  (\bibinfo{year}{2013}).

\bibitem[{\citenamefont{Amelino-Camelia}(2002)}]{amelino2002relativity}
\bibinfo{author}{\bibfnamefont{G.}~\bibnamefont{Amelino-Camelia}},
  \bibinfo{journal}{International Journal of Modern Physics D}
  \textbf{\bibinfo{volume}{11}}, \bibinfo{pages}{35} (\bibinfo{year}{2002}).

\end{thebibliography}

\end{document}